\documentclass[]{aa}
\usepackage{graphicx}
\usepackage{txfonts}
\usepackage{aalongtable}
\usepackage{natbib}
\bibpunct{(}{)}{;}{a}{}{,}
\begin{document}
\title{$\beta$~Cephei stars in the ASAS-3 data}
\subtitle{II. 103 new $\beta$~Cephei stars and a discussion of low-frequency modes}
\author{A.\,Pigulski\inst{1} \and G.\,Pojma\'nski\inst{2}}
\offprints{A.\,Pigulski}
\institute{Instytut Astronomiczny Uniwersytetu Wroc{\l}awskiego,
Kopernika 11, 51-622 Wroc{\l}aw, Poland\\
\email{ pigulski@astro.uni.wroc.pl} \and
Obserwatorium Astronomiczne Uniwersytetu Warszawskiego,
Al.~Ujazdowskie 4, 00-478 Warszawa\\
\email{gp@astrouw.edu.pl}}
\date{Received August 30, 2007/ Accepted November 7, 2007}
\abstract{The $\beta$~Cephei stars have been studied for over a hundred years. Despite this, many interesting problems related to this class
of variable stars remain unsolved.  Fortunately, these stars seem to be well-suited to asteroseismology. Hence, the results of seismic analysis of
$\beta$~Cephei stars should help us to better understand pulsations and the main sequence evolution of massive stars, particularly
the effect of rotation on mode excitation and internal structure. It is therefore extremely important to increase the sample of known
$\beta$~Cephei stars and select targets that are useful for asteroseismology.}
{We analysed ASAS-3 photometry of bright early-type stars with the goal of finding new $\beta$~Cephei stars. We were particularly interested in 
$\beta$~Cephei stars that would be good for seismic analysis, 
i.e., stars that (i) have a large number of excited modes, (ii) show rotationally split modes,
(iii) are components of eclipsing binary systems, (iv) have low-frequency modes, that is, are hybrid $\beta$~Cephei/SPB stars.}
{Our study was made with a homogeneous sample of over 4100 stars having MK spectral type B5 or earlier. For these
stars, the ASAS-3 photometry was analysed by means of a Fourier periodogram.}
{We have discovered 103 $\beta$~Cephei stars, nearly doubling the number of previously known stars of this type. Among these stars, four are components of eclipsing binaries, seven have modes equidistant or nearly equidistant in frequency.
In addition, we found five $\beta$~Cephei stars that show low-frequency periodic variations, very likely due to pulsations. We therefore regard them as candidate hybrid
$\beta$~Cephei/SPB pulsators. All these stars are potentially very useful for seismic modeling. Moreover, we found $\beta$~Cephei-type pulsations in
three late O-type stars and fast period changes in one, HD\,168050.}
{}

\maketitle

\section{Introduction}
This is the second paper in the series presenting results of a search for $\beta$~Cephei-type pulsating variables in the
$V$-filter photometry in the third phase of the All Sky Automated Survey, ASAS-3 \citep{pojm97,pojm00,pojm05}.
In the first paper of the series \citep[][hereafter Paper I]{pipo07}, we published the results of an analysis of the ASAS-3 photometry for 41
of about 112 presently known $\beta$~Cephei-type stars, that is, all that had sufficiently good photometry in the ASAS-3 database.
The analysis allowed us to detect amplitude changes in three $\beta$~Cephei stars, BW~Cru, V836~Cen and V348~Nor, and period changes
in KK~Vel and V836~Cen. In addition, new modes were found or ambiguities in the frequencies of previously detected modes were removed for some stars.

The published ASAS catalogues, containing photometry of variable star candidates, have already been used to discover $\beta$~Cephei stars.
In total, 19 new members of this group have been found in the ASAS data \citep{pigu05,hand05}. However, owing to the fact that the method
of selection of variable stars was biased towards large-amplitude variables, all these 19 stars had relatively large amplitudes.
Therefore, it was obvious that once a detailed analysis of the whole ASAS-3 database has been done, many new stars
of this type, with smaller amplitudes, will be found. Such an analysis will be carried out in the future, but because the whole ASAS database contains millions
of stars and the period(s) alone are not sufficient to classify a given star as a $\beta$~Cephei variable, we decided to proceed in a different way.
Using the information in the existing stellar catalogues, we selected stars that are known to be either O-type or early B-type stars and
analysed ASAS-3 photometry for these stars only. The selection criteria are explained in Sect.~2. As a result, nearly 300 $\beta$~Cephei stars were discovered of which the first part is discussed in this paper.  This will allow us to define the properties of the whole class much better and,
in view of the increasing interest in the study of these stars by means of asteroseismology, to select appropriate targets for subsequent work.
In the present paper we describe the results of the search for $\beta$~Cephei-type stars among objects selected
from the five volumes of the Michigan Catalogue for HD stars.  The ASAS-3 photometry for stars selected using other catalogues
will be discussed in the next paper.

\section{Selection of early-type stars}
The existing catalogues contain many different types of data that allow selecting early-type stars. The early B and O-type stars can
be best selected using either the information on spectral types or the $UBV$ photometry. For this reason, the obvious choice
of the source of the information on spectral types was the {\it Michigan Catalogue of Two-Dimensional Spectral Types for the HD stars} 
\citep{houk1,houk2,houk3,houk4,houk5}.
The five volumes of this catalogue contain MK spectral types for 161\,472 stars south of declination $+$5$\degr$. The stars are the same as those
in {\it The Henry Draper Catalogue} \citep{hd}, i.e., are brighter than $\sim$10 mag in $V$.
This means that all stars fainter than $\sim$7 mag with HD numbers will have good ASAS-3 photometry because the ASAS survey covers
the range of $V$ magnitudes between 7 and 14~mag.
\begin{figure*}
\centering
\includegraphics[width=15cm]{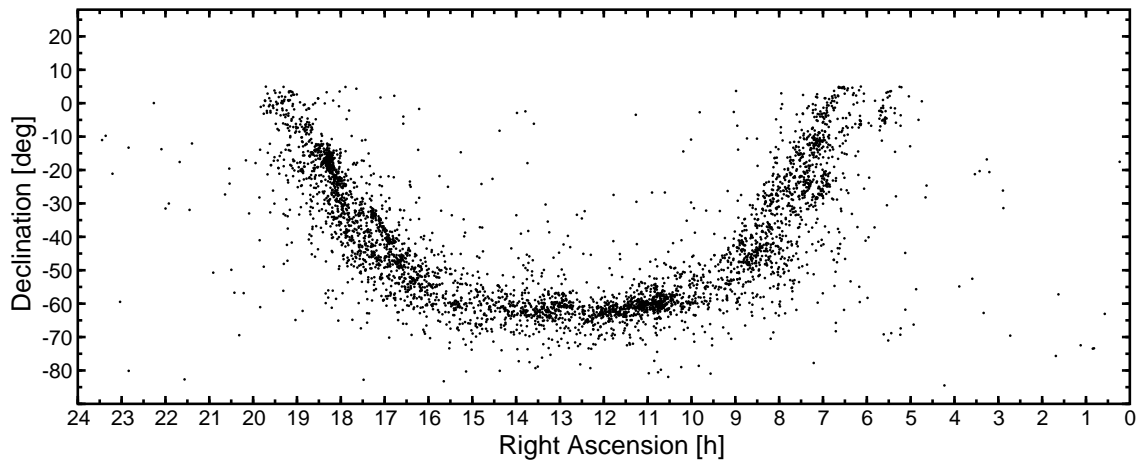}
\caption{Location in equatorial coordinates of all stars with $V > $ 7 mag, $\delta < +$5$\degr$ and of spectral type B5 or earlier, selected from Michigan Catalogues
as candidates for $\beta$~Cephei stars.}
\label{houk-B-stars}
\end{figure*}

While selecting stars from the Michigan Catalogue, we used only two conditions: (i) $V >$ 7~mag (brighter stars are overexposed
in the ASAS $V$-filter frames), (ii) MK spectral type B5 or earlier.
In this way, about 4100 stars were selected.  Their equatorial coordinates are plotted in Fig.~\ref{houk-B-stars}. As expected,
the stars concentrate along the Galactic plane. For all these stars, the ASAS-3 photometry was extracted, and subsequently analysed
by means of Fourier periodograms in the frequency range from 0 to 40~d$^{-1}$.

A star was regarded as variable if the signal-to-noise (S/N) of the highest peak in the periodogram exceeded 5.

\section{The results}
In total, 103 $\beta$~Cephei stars were
found in the sample of about 4100 stars selected using Michigan Catalogue. These stars are listed in Tab.~\ref{bc-new};
the frequencies, amplitudes and times of maximum light for the detected modes are presented in Tab.~\ref{bc-new-f}, available online. In addition,
periodograms showing the consecutive steps of prewhitening are shown in Figs.~\ref{fp-01}--\ref{fp-13}, also available as online material.\footnote{The $V$ photometry for all 103 stars is available in electronic form at the CDS via anonymous ftp to
cdsarc.u-strasbg.fr (130.79.128.5) or via http://cdsweb.u-strasbg.fr/cgi-bin/qcat?J/A+A/??/??}

\addtocounter{table}{1}

\onltab{2}{%
{\small
\begin{longtable}{rcclrlrr}
\caption{\label{bc-new-f}Parameters of the sine-curve fits to the $V$ magnitudes of new $\beta$~Cephei-type stars listed in Tab.~\ref{bc-new}.
$N_{\rm obs}$ is the number of observations, the initial epoch $T_0$ is equal to 2450000.0.  The other parameters are the following:
$T_{\rm max}^i$, time of maximum light for the $i$th mode, $\sigma_{\rm res}$, standard deviation of the residuals, DT, detection threshold.}\\
\hline\hline
  & & & \multicolumn{1}{c}{$f_{\rm i}$} & \multicolumn{1}{c}{$A_i$} & \multicolumn{1}{c}{$T_{\rm max}^i - T_0$} & 
\multicolumn{1}{c}{$\sigma_{\rm res}$} & \multicolumn{1}{c}{DT}\\
Star name &  Freq. & $N_{\rm obs}$ & \multicolumn{1}{c}{[d$^{-1}$]} & \multicolumn{1}{c}{[mmag]} & \multicolumn{1}{c}{[d]}
& \multicolumn{1}{c}{[mmag]} & \multicolumn{1}{c}{[mmag]} \\
\hline
\endfirsthead
\caption{continued.}\\
\hline
\endhead
\hline
\endfoot
\object{HD 46994} & $f_1$ & 322 & 3.90041(4) & 7.8(10) & 2942.1701(54) & 12.8 & 5.1\\
\object{HD 48553} & $f_1$ & 215 & 5.59806(4) & 9.1(10) & 3143.6502(31) & 10.2 & 4.9\\
\object{HD 67600} & $f_1$ & 375 & 3.78661(2) & 8.8(08) & 2901.9390(35) & 10.2 & 3.8\\
\object{HD 68962} & $f_1$ & 191 & 4.76251(6) & 7.4(09) & 3069.0758(38) & 8.4 & 4.3\\
\object{HD 69016} & $f_1$ & 616 & 5.01186(4) & 6.4(09) & 2816.8526(46) & 15.3 & 4.4\\
\object{HD 69824} & $f_1$ & 351 & 6.06354(4) & 5.2(07) & 2826.1460(37) & 9.5 & 3.6\\
\object{HD 73568} & $f_1$ & 576 &  4.54531(3) & 4.8(06) & 2865.8111(42) & 9.7 & 2.9\\
& $f_2$ & & 6.58022(5) & 3.4(06) & 2865.6858(41) & & \\
\object{HD 74339} & $f_1$ & 432 & 5.220103(10) & 27.7(09) & 2900.6750(10) & 13.1 & 4.5\\
& $f_2$ & & 5.195216(13) & 20.7(09) &  2900.7085(14) & &\\
& $f_3$ & & 5.22221(2) & 13.7(09) & 2900.6380(21) & & \\
& $f_4$ & & 5.23894(3) & 11.5(09) & 2900.6309(25) & & \\
& $f_5$ & & 5.20531(3) & 10.6(09) & 2900.6775(26) & & \\
& $f_6$ & & 5.25563(3) & 10.3(09) & 2900.6234(27) & & \\
\object{HD 77769} & $f_1$ & 373 & 5.46831(4) & 5.5(08) & 2849.0965(46) & 11.5 & 4.2\\
\object{HD 86085} & $f_1$ & 379 & 7.87529(2) & 9.2(08) & 2840.1421(17) & 10.5 & 3.9\\
\object{HD 86214} & $f_1$ & 512 & 4.442908(13) & 14.3(08) & 2958.9401(20) & 11.5 & 3.6\\
& $f_2$ & & 4.40462(2) & 10.9(07) & 2959.0537(24) & &\\
& $f_3$ & & 4.40967(3) & 8.3(07) & 2958.9773(32) & &\\
& $f_4$ & & 4.41462(3) & 9.0(07) & 2958.9914(29) & &\\
& $f_5$ & & 4.39994(4) & 6.3(08) & 2958.8671(42) & &\\
& $f_6$ & & 4.44618(4) & 5.7(08) & 2959.0029(48) & &\\
\object{HD 86248} & $f_1$ & 395 & 8.25590(5) & 5.5(08) & 2835.6249(29) & 11.4 & 4.2\\
\object{HD 87592} & $f_1$ & 790 & 5.15127(2) & 6.9(06) & 2828.1611(26) & 11.3 & 2.8\\
& $f_2$ & &  5.09753(3) & 6.8(06) & 2828.1263(26) & &\\
& $f_3$ & & 5.43571(3) & 6.3(06) & 2828.1232(27) & &\\
& $f_4$ & & 5.30697(4) & 4.1(06) & 2828.1605(43) & &\\
& $f_5$ & & 5.33044(6) & 2.9(06) & 2828.1757(60) & &\\
\object{HD 88844} & $f_1$ & 373 & 5.37589(3) & 7.5(07) & 2881.4701(27) & 9.6 & 3.5\\
& $f_2$ & & 6.03797(3) & 5.9(07) & 2881.4999(31) & &\\
\object{HD 90075} & $f_1$ & 371 & 4.09526(3) & 10.3(09) & 2902.7893(34) & 12.2 & 4.5\\
\object{HD 90987} & $f_1$ & 368 & 4.702781(07) & 39.6(10) & 2873.8688(08) & 13.2 & 4.9\\
& $f_2$ & & 4.56652(2) & 16.1(10) & 2874.0208(21) & &\\
& $f_3$ & & 4.59644(3) & 8.7(10) & 2873.8866(39) & &\\
& $f_4$ & & 4.36623(4) & 6.7(10) & 2873.9638(54) & &\\
\object{HD 91651} & $f_1$ & 525 & 7.38876(3) & 5.6(06) & 2876.1314(23) & 9.7 & 3.0\\
& $f_2$ & & 7.10851(5) & 3.5(06) & 2876.1737(38) & &\\
\object{HD 92291} & $f_1$ & 385 & 5.528817(10) & 23.3(09) & 2884.2608(11) & 11.8 & 4.2\\
& $f_2$ & & 5.454284(13) & 17.5(09) & 2884.2939(15) & &\\
& $f_3$ & & 5.90508(3) & 8.9(09) & 2884.3169(26) & &\\
& $f_4$ & & 5.46980(3) & 7.4(09) & 2884.2436(34) & &\\
& $f_5$ & & 5.72446(4) & 6.2(08) & 2884.2426(40) & &\\
\object{HD 93113} & $f_1$ & 690 & 5.57541(5) & 3.7(06) & 2856.9991(46) & 11.2 & 3.1\\
\object{HD 93341} & $f_1$ & 691 & 5.08749(3) & 8.2(07) & 2881.2753(24) & 12.1 & 3.3\\
\object{HD 94065} & $f_1$ & 364 & 4.810101(10) & 20.3(08) & 2861.4725(13) & 10.3 & 3.8\\
& $f_2$ & & 5.02213(3) & 11.1(08) & 2861.4386(24) & &\\
& $f_3$ & & 5.01919(2) & 13.5(08) & 2861.5756(19) & &\\
& $f_4$ & & 5.48660(3) & 9.3(08) & 2861.4498(24) & &\\
& $f_5$ & & 5.02074(3) & 8.5(08) & 2861.4767(30) & &\\
\object{HD 94345} & $f_1$ & 487 & 5.16834(2) & 9.4(07) & 2994.2366(23) & 10.6 & 3.4\\
& $f_2$ & & 5.17063(4) & 5.4(07) & 2994.3449(39) & &\\
& $f_3$ & & 5.57486(5) & 3.8(07) & 2994.2818(53) & &\\
\object{HD 94900} & $f_1$ & 468 & 4.842766(11) & 16.8(08) & 2930.6143(14) & 11.2 & 3.7\\
\object{HD 95568} & $f_1$ & 365 & 6.152355(15) & 16.9(10) & 2857.4896(15) & 12.7 & 4.7\\
& $f_2$ & & 6.13549(3) & 10.4(10) & 2857.4369(24) & &\\
\object{HD 96882} & $f_1$ & 355 & 6.22739(4) & 6.6(09) & 2909.9008(33) & 11.3 & 4.3\\
\object{HD 96901} & $f_1$ & 340 & 4.163209(11) & 19.2(08) & 2900.2036(16) & 10.1 & 3.9\\
& $f_2$ & & 4.128028(17) & 11.4(08) & 2900.2624(27) & &\\
\object{HD 97629} & $f_1$ & 366 & 5.178579(17) & 13.3(08) & 2920.9323(19) & 11.0 & 4.1\\
\object{HD 98260} & $f_1$ & 350 & 4.514510(19) & 14.2(09) & 2896.2087(23) & 12.0 & 4.6\\
& $f_2$ & & 4.44349(3) & 9.8(09) & 2896.1907(34) & &\\
& $f_3$ & & 4.48065(3) & 10.3(09) & 2896.0648(31) & &\\
\object{HD 99024} & $f_1$ & 406 & 7.03125(4) & 7.0(08) & 2992.2193(28) & 12.0 & 4.2\\
\object{HD 99205} & $f_1$ & 645 & 3.822544(10) & 18.1(07) & 2882.5932(16) & 12.3 & 3.5\\
& $f_2$ & & 1.24684(4) & 5.7(07) & 2882.334(16) & &\\
\object{HD 100355} & $f_1$ & 670 & 7.507394(15) & 10.3(06) & 2865.7310(13) & 10.9 & 3.0\\
& $f_2$ & & 7.112278(16) & 10.2(06) & 2865.7452(13) & &\\
& $f_3$ & & 6.93375(4) & 4.2(06) & 2865.6535(32) & &\\
\object{HD 101794} & $f_1$ & 311 & 4.45140(3) & 14.6(13) & 2737.1557(31) & 15.4 & 6.2\\
& $f_2$ & & 1.83951(4) & 11.2(13) & 2736.998(10) & &\\
\object{HD 101838} & $f_1$ & 352 & 3.12760(3) & 10.3(09) & 2865.5247(45) & 11.8 & 4.9\\
\object{HD 102505} & $f_1$ & 332 & 4.750770(15) & 13.7(08) & 2865.3997(19) & 9.9 & 3.9\\
& $f_2$ & & 4.52947(4) & 5.8(08) & 2865.4186(47) & &\\
\object{HD 103007} & $f_1$ & 346 & 5.707239(18) & 12.2(09) & 2849.9655(20) & 11.2 & 4.2\\
& $f_2$ & & 5.75454(3) & 10.5(09) & 2849.9456(23) & &\\
& $f_3$ & & 5.67467(3) & 9.9(09) & 2849.9219(25) & &\\
\object{HD 103320} & $f_1$ & 351 & 4.71222(2) & 8.0(08) & 2894.0787(32) & 9.8 & 3.7\\
\object{HD 103764} & $f_1$ & 330 & 5.582626(18) & 12.2(09) & 2859.0861(20) & 10.9 & 4.3\\
\object{HD 104257} & $f_1$ & 321 & 6.868277(11) & 20.6(09) & 2852.5829(10) & 11.0 & 4.4\\
\object{HD 104465} & $f_1$ & 330 & 5.518702(17) & 14.8(10) & 2862.6886(19) & 12.2 & 4.8\\
& $f_2$ & & 5.395199(19) & 14.2(10) & 2862.7541(20) & &\\
\object{HD 104795} & $f_1$ & 330 & 6.23215(4) & 5.3(07) & 2883.9586(31) & 8.4 & 3.3\\
\object{HD 106345} & $f_1$ & 440 & 6.612477(12) & 15.0(07) & 2843.0066(11) & 10.2 & 3.4\\
\object{HD 108628} & $f_1$ & 637 & 6.74262(4) & 4.9(06) & 2867.9268(30) & 10.9 & 3.1\\
\object{HD 108769} & $f_1$ & 317 & 7.54085(4) & 5.4(08) & 2816.7267(32) & 10.1 & 4.0\\
\object{HD 110498} & $f_1$ & 325 & 4.15383(2) & 10.3(09) & 2875.1748(30) & 10.5 & 4.1\\
\object{HD 111377} & $f_1$ & 342 & 8.83122(4) & 6.4(09) & 2933.8111(25) & 11.4 & 4.4\\
\object{HD 111578} & $f_1$ & 344 & 4.71676(3) & 8.8(09) & 2914.2855(34) & 11.7 & 4.5\\
\object{HD 113013} & $f_1$ & 393 & 6.47119(3) & 9.5(08) & 2986.8716(22) & 11.7 & 4.2\\
& $f_2$ & & 6.26890(4) & 6.4(08) & 2986.8920(34) & &\\
\object{HD 114444} & $f_1$ & 508 & 4.81618(4) & 8.0(10) & 2709.0849(43) & 16.3 & 5.1\\
& $f_2$ & & 5.11817(5) & 6.6(10) & 2709.1424(49) & &\\
\object{HD 114733} & $f_1$ & 620 & 7.24916(3) & 8.3(07) & 2854.8118(19) & 12.4 & 3.5\\
\object{HD 115533} & $f_1$ & 330 & 14.57715(3) & 11.8(11) & 2869.9716(10) & 14.0 & 5.5\\
\object{HD 116538} & $f_1$ & 365 & 4.52246(3) & 10.6(11) & 2859.5868(36) & 14.4 & 5.4\\
\object{HD 116827} & $f_1$ & 555 & 2.21158(3) & 12.5(09) & 2968.2039(52) & 15.0 & 4.5\\
& $f_2$ & & 4.45852(3) & 10.4(09) & 2967.9669(31) & &\\
\object{HD 117357} & $f_1$ & 561 & 2.08106(3) & 8.8(07) & 2967.5187(57) & 11.0 & 3.3\\
& $f_2$ & & 6.53990(4) & 6.2(07) & 2967.7211(26) & &\\
\object{HD 117687} & $f_1$ & 515 & 6.17482(3) & 6.8(07) & 2944.0268(25) & 10.6 & 3.3\\
\object{HD 117704} & $f_1$ & 559 & 4.65537(3) & 8.4(08) & 2971.8997(34) & 13.8 & 4.2\\
\object{HD 119252} & $f_1$ & 326 & 8.78623(3) & 13.3(11) & 2863.9353(14) & 13.5 & 5.3\\
\object{HD 119910} & $f_1$ & 314 & 7.03488(5) & 6.9(07) & 3161.7560(24) & 9.2 & 3.7\\
& $f_2$ & & 5.95546(6) & 4.6(07) & 3161.7249(43) & &\\
\object{HD 122831} & $f_1$ & 462 & 5.008025(17) & 14.9(09) & 3023.4327(19) & 12.7 & 4.1\\
& $f_2$ & & 1.82715(5) & 4.9(08) & 3023.337(15) & &\\
\object{HD 123077} & $f_1$ & 444 & 7.043592(14) & 23.6(11) & 2811.6284(10) & 15.7 & 5.3\\
& $f_2$ & & 6.95371(3) & 10.2(10) & 2811.6703(25) & &\\
& $f_3$ & & 6.62910(4) & 9.8(11) & 2811.5793(27) & &\\
& $f_4$ & & 7.70277(6) & 6.4(11) & 2811.5844(35) & &\\
\object{HD 126357} & $f_1$ & 516 & 4.06927(3) & 8.4(09) & 2754.2463(38) & 13.1 & 4.1\\
& $f_2$ & & 8.54271(4) & 8.1(08) & 2754.1686(19) & &\\
& $f_3$ & & 8.07010(4) & 7.8(08) & 2754.0913(21) & &\\
\object{HD 131805} & $f_1$ & 546 & 7.38299(4) & 5.9(08) & 2808.2366(28) & 12.8 & 3.9\\
\object{HD 132320} & $f_1$ & 377 & 8.44635(3) & 7.6(09) & 2808.8232(22) & 12.1 & 4.4\\
\object{HD 137405} & $f_1$ & 677 & 4.334276(15) & 11.3(06) & 2924.5163(19) & 10.9 & 2.9\\
\object{HD 142754} & $f_1$ & 289 & 8.51115(5) & 6.8(10) & 2880.1249(28) & 11.9 & 5.0\\
\object{HD 145537} & $f_1$ & 284 & 5.481110(17) & 22.9(13) & 2849.0793(17) & 15.3 & 6.5\\
& $f_2$ & & 5.18768(3) & 16.4(13) & 2849.0567(24) & &\\
\object{HD 146442} & $f_1$ & 328 & 6.57248(3) & 9.1(09) & 2899.1253(22) & 10.7 & 4.3\\
& $f_2$ & & 6.76650(4) & 6.0(09) & 2899.0557(33) & &\\
& $f_3$ & & 5.50073(5) & 5.4(08) & 2899.0582(46) & &\\
\object{HD 147421} & $f_1$ & 642 & 5.354726(06) & 27.1(06) & 2899.7481(06) & 9.9 & 2.8\\
\object{HD 149100} & $f_1$ & 284 & 5.16118(4) & 8.1(08) & 3128.9118(30) & 9.2 & 3.8\\
\object{HD 150927} & $f_1$ & 328 & 4.834676(15) & 24.7(11) & 2916.1050(16) & 14.5 & 5.7\\
& $f_2$ & & 4.854070(16) & 22.4(12) & 2916.1120(17) & &\\
\object{HD 151158} & $f_1$ & 307 & 5.501130(17) & 17.5(10) & 2892.7554(17) & 12.4 & 5.3\\
\object{HD 152060} & $f_1$ & 303 & 9.57323(3) & 8.5(08) & 2884.4547(17) & 10.3 & 4.2\\
& $f_2$ & & 12.19261(3) & 8.8(08) & 2884.4692(12) & &\\
\object{HD 152162} & $f_1$ & 490 & 6.031971(06) & 33.1(06) & 2867.8221(05) & 9.7 & 3.3\\
& $f_2$ & & 6.02169(2) & 8.6(06) & 2867.7874(19) & &\\
\object{HD 152372} & $f_1$ & 479 & 8.64280(3) & 8.6(07) & 2840.2601(15) & 10.6 & 3.5\\
& $f_2$ & & 8.33283(4) & 5.7(07) & 2840.2841(23) & &\\
& $f_3$ & & 8.68004(5) & 4.4(07) & 2840.3162(28) & &\\
\object{HD 153772} & $f_1$ & 351 & 3.275788(10) & 23.9(08) & 2874.3683(16) & 10.8 & 4.1\\
\object{HD 154500} & $f_1$ & 911 & 6.98730(3) & 8.6(06) & 2991.7359(17) & 13.1 & 3.6\\
& $f_2$ & & 7.05945(3) & 7.9(06) & 2991.8199(18) & &\\
\object{HD 155407} & $f_1$ & 319 & 5.552652(09) & 28.9(09) & 2865.0140(09) & 11.2 & 4.4\\
\object{HD 156172} & $f_1$ & 294 & 6.97258(3) & 10.8(10) & 2909.6293(21) & 12.4 & 5.1\\
\object{HD 156321} & $f_1$ & 631 & 8.72485(5) & 4.0(05) & 3098.6986(22) & 8.3 & 2.5\\
\object{HD 159792} & $f_1$ & 536 & 7.000049(10) & 25.7(09) & 2801.4000(08) & 14.7 & 4.5\\
& $f_2$ & & 7.24311(3) & 8.4(09) & 2801.3910(24) & &\\
& $f_3$ & & 6.91847(5) & 5.6(09) & 2801.3297(37) & &\\
\object{HD 161633} & $f_1$ & 332 & 6.10395(2) & 12.8(10) & 2828.7628(21) & 12.9 & 5.0\\
& $f_2$ & & 6.26392(3) & 11.4(10) & 2828.6801(23) & &\\
& $f_3$ & & 5.90006(3) & 10.6(10) & 2828.6718(26) & &\\
& $f_4$ & & 5.60499(5) & 5.9(10) & 2828.7392(49) & &\\
\object{HD 164188} & $f_1$ & 413 & 5.71220(4) & 5.8(07) & 2903.4761(30) & 9.3 & 3.3\\
\object{HD 164741} & $f_1$ & 474 & 5.119632(19) & 13.9(08) & 2846.7981(19) & 12.0 & 3.9\\
& $f_2$ & & 5.16521(2) & 12.0(08) & 2846.8063(20) & &\\
& $f_3$ & & 5.56809(3) & 8.6(08) & 2846.8316(27) & &\\
& $f_4$ & & 5.12946(4) & 6.6(08) & 2846.6758(38) & &\\
& $f_5$ & & 4.96110(5) & 5.1(08) & 2846.7909(51) & &\\
\object{HD 165955} & $f_1$ & 762 & 11.906129(19) & 11.0(06) & 3383.4615(07) & 10.9 & 3.2\\
& $f_2$ & & 10.16581(4) & 7.4(06) & 3383.4489(12) & &\\
\object{HD 166304} & $f_1$ & 633 & 7.03862(3) & 5.9(06) & 2888.9596(20) & 9.3 & 2.7\\
& $f_2$ & & 6.33937(6) & 3.3(06) & 2888.9669(39) & &\\
& $f_3$ & & 7.55353(6) & 3.3(06) & 2888.8784(33) & &\\
\object{HD 167003} & $f_1$ & 758 & 6.77267(3) & 10.5(06) & 3395.4748(13) & 10.8 & 3.2\\
& $f_2$ & & 7.54580(4) & 6.8(06) & 3395.5124(18) & &\\
& $f_3$ & & 7.01603(4) & 6.2(06) & 3395.4349(21) & &\\
& $f_4$ & & 5.37835(4) & 5.7(06) & 3395.5481(29) & &\\
\object{HD 167451} & $f_1$ & 330 & 4.37474(3) & 11.5(09) & 2849.3065(29) & 11.2 & 4.4\\
\object{HD 168015} & $f_1$ & 349 & 6.324050(17) & 12.4(07) & 2830.9715(14) & 9.0 & 3.4\\
& $f_2$ & & 5.07961(2) & 10.1(07) & 2831.0446(21) & &\\
\object{HD 168050} & $f_1$ & 698 & 5.5486... [var] & 27.7(10) & see Tab.~\ref{168050tmax} & 12.2 & 3.7\\
& $f_2$ & & 5.251039(15) & 20.7(10) & 2921.1561(15) & &\\
\object{HD 168675} & $f_1$ & 596 & 6.35906(2) & 8.7(06) & 2867.6437(17) & 9.7 & 2.8\\
& $f_2$ & & 4.18134(4) & 5.1(06) & 2867.6660(42) & &\\
\object{HD 168750} & $f_1$ & 547 & 4.061855(12) & 15.5(06) & 2849.0681(16) & 10.2 & 3.1\\
& $f_2$ & & 4.025154(19) & 9.5(06) & 2849.1496(26) & &\\
& $f_3$ & & 4.04348(4) & 5.1(06) & 2849.2132(49) & &\\
\object{HD 169173} & $f_1$ & 798 & 5.31303(3) & 6.7(07) & 2883.5407(29) & 13.0 & 3.6\\
\object{HD 169601} & $f_1$ & 612 & 5.173682(11) & 23.0(08) & 2844.3032(10) & 12.3 & 3.9\\
& $f_2$ & & 5.231036(18) & 14.2(07) & 2844.3146(16) & &\\
& $f_3$ & & 5.08657(3) & 8.2(07) & 2844.2888(28) & &\\
\object{HD 171141} & $f_1$ & 349 & 11.19358(4) & 7.1(07) & 2845.9130(14) & 9.0 & 3.5\\
\object{HD 171305} & $f_1$ & 288 & 5.15860(2) & 10.0(08) & 2937.3750(24) & 9.3 & 3.9\\
\object{HD 171344} & $f_1$ & 316 & 5.41534(5) & 7.5(10) & 2828.6084(37) & 11.8 & 4.7\\
\object{HD 172140} & $f_1$ & 1346 & 6.61531(3) & 8.8(04) & 2867.4248(11) & 9.5 & 2.1\\
& $f_2$ & & 6.29880(3) & 7.8(04) & 2867.3434(14) & &\\
& $f_3$ & & 7.17605(4) & 6.2(04) & 2867.4152(14) & &\\
& $f_4$ & & 6.25128(6) & 4.2(04) & 2867.4350(25) & &\\
& $f_5$ & & 7.56117(9) & 2.6(04) & 2867.4188(30) & &\\
\object{HD 172427} & $f_1$ & 283 & 4.01969(3) & 10.6(09) & 2840.9159(32) & 9.9 & 4.2\\
\object{HD 173006} & $f_1$ & 269 & 5.878262(06) & 87.8(18) & 2892.8486(05) & 20.7 & 9.0\\
& 2$f_1$ & & 11.756524 & 22.9(18) & 2892.7824(11) & &\\
\object{HD 173502} & $f_1$ & 1342 & 5.496527(07) & 37.3(04) & 2865.1104(04) & 9.5 & 2.2\\
& $f_2$ & & 5.540236(15) & 16.4(04) & 2865.0533(07) & &\\
& $f_3$ & & 5.33345(3) & 8.1(04) & 2865.0716(15) & &\\
& $f_4$ & & 5.15802(8) & 3.2(04) & 2865.0158(37) & &\\
\object{HD 178987} & $f_1$ & 339 & 7.14660(3) & 10.5(11) & 2854.0577(23) & 14.4 & 5.7\\
\object{HD 180032} & $f_1$ & 287 & 6.86806(3) & 8.6(08) & 2820.2828(21) & 9.3 & 3.9\\
& $f_2$ & & 7.18637(4) & 5.9(08) & 2820.4152(29) & &\\
\object{HD 186610} & $f_1$ & 726 & 5.499983(17) & 15.6(06) & 3023.3878(11) & 10.8 & 2.9\\
& $f_2$ & & 5.76342(3) & 13.0(06) & 3023.3979(12) & &\\
& $f_3$ & & 5.46127(3) & 11.4(06) & 3023.4022(15) & &\\
& $f_4$ & & 5.53861(5) & 5.3(06) & 3023.5457(31) & &\\
\object{HD 187536} & $f_1$ & 285 & 4.58572(4) & 6.7(09) & 2796.7750(46) & 10.7 & 4.5\\
\hline
\end{longtable}
}
}

\onlfig{2}{%
\begin{figure*}
\centering
\includegraphics[width=17cm]{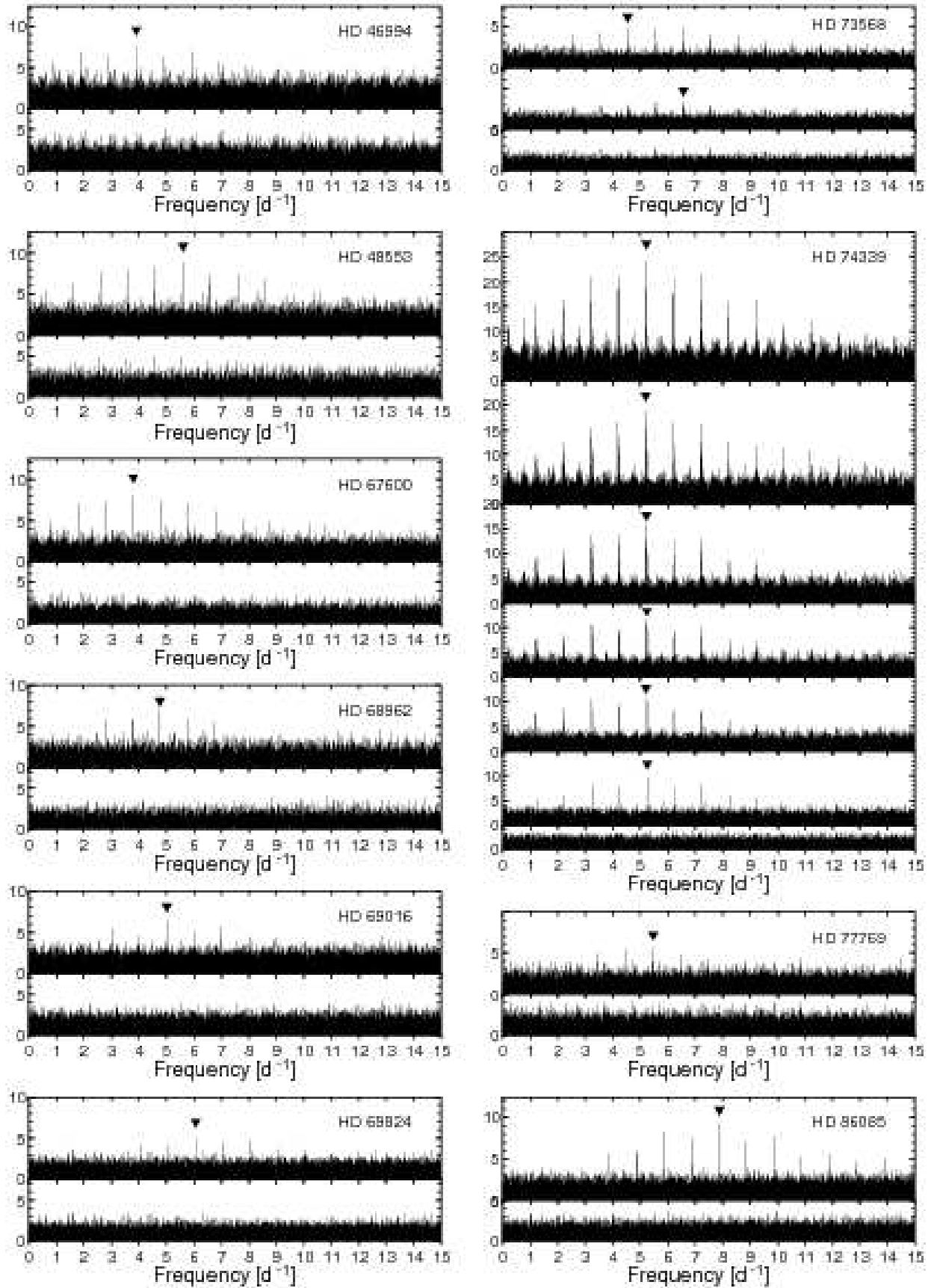}
\caption{Fourier periodograms of the ASAS-3 data of ten $\beta$~Cephei-type stars discussed in the paper: HD\,46994, 48553, 67600, 68962, 69016, 69824,
73568, 74339, 77769, and 86085. Lower panels show consecutive steps of prewhitening.  Frequencies of the detected modes are
indicated by inverted triangles.  Ordinate is the amplitude expressed in mmag.}
\label{fp-01}
\end{figure*}
}

\onlfig{3}{%
\begin{figure*}
\centering
\includegraphics[width=17cm]{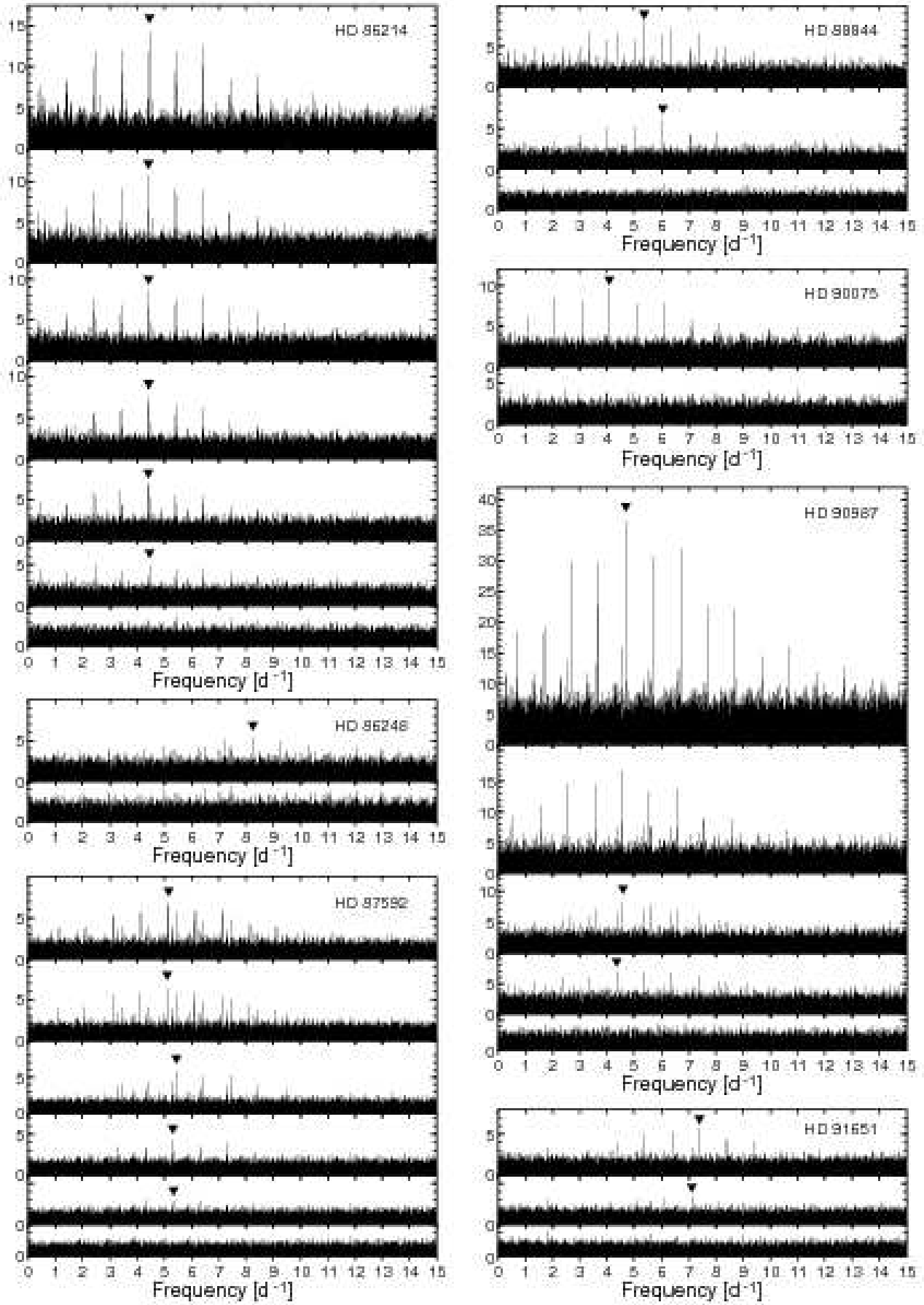}
\caption{The same as in Fig.~\ref{fp-01}, but for HD 86214, 86248, 87592, 88844, 90075, 90987, and 91651.}
\label{fp-02}
\end{figure*}
}

\onlfig{4}{%
\begin{figure*}
\centering
\includegraphics[width=17cm]{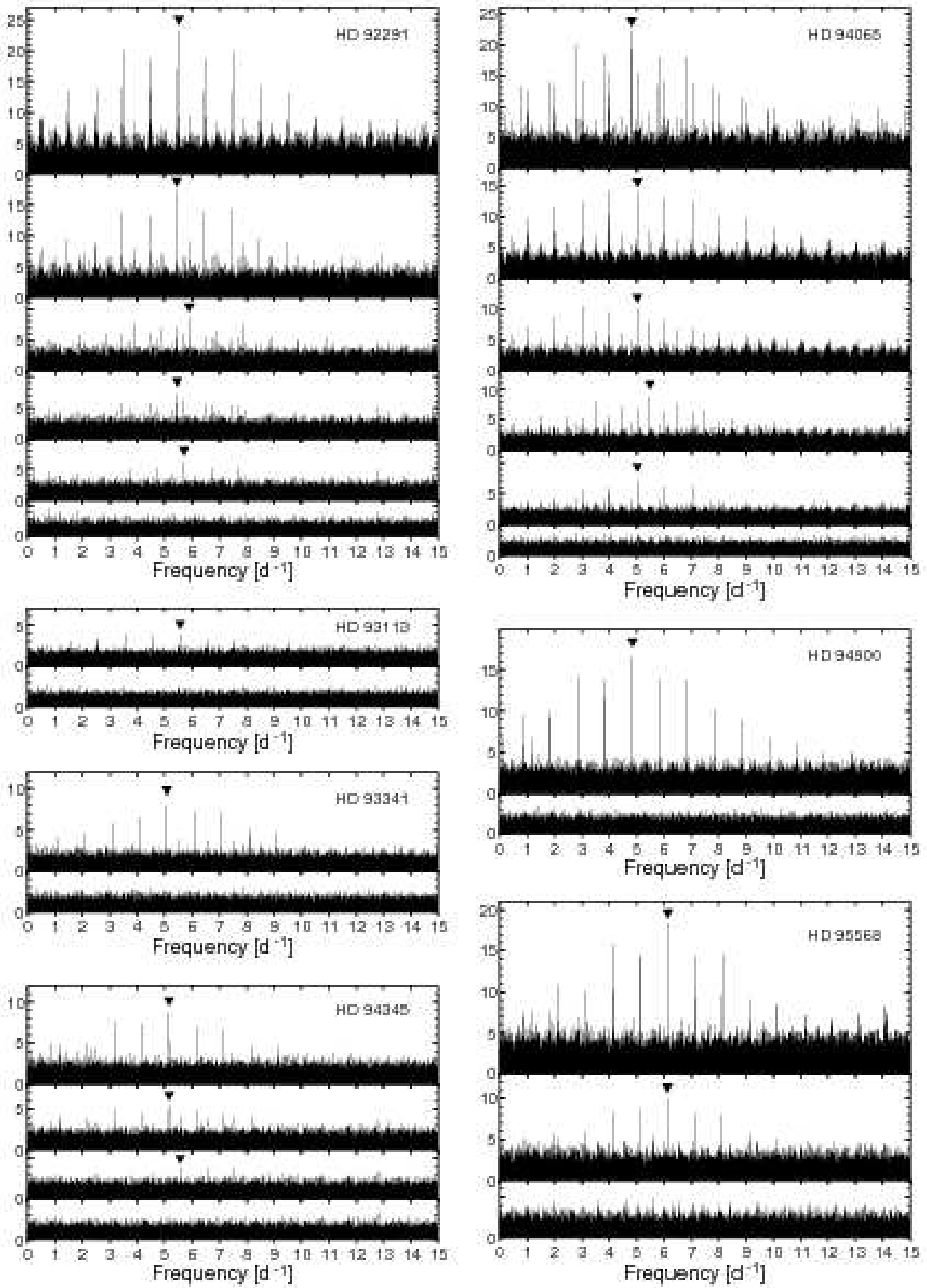}
\caption{The same as in Fig.~\ref{fp-01}, but for HD 92291, 93113, 93341, 94065, 94345, 94900, and 95568.}
\label{fp-03}
\end{figure*}
}

\onlfig{5}{%
\begin{figure*}
\centering
\includegraphics[width=17cm]{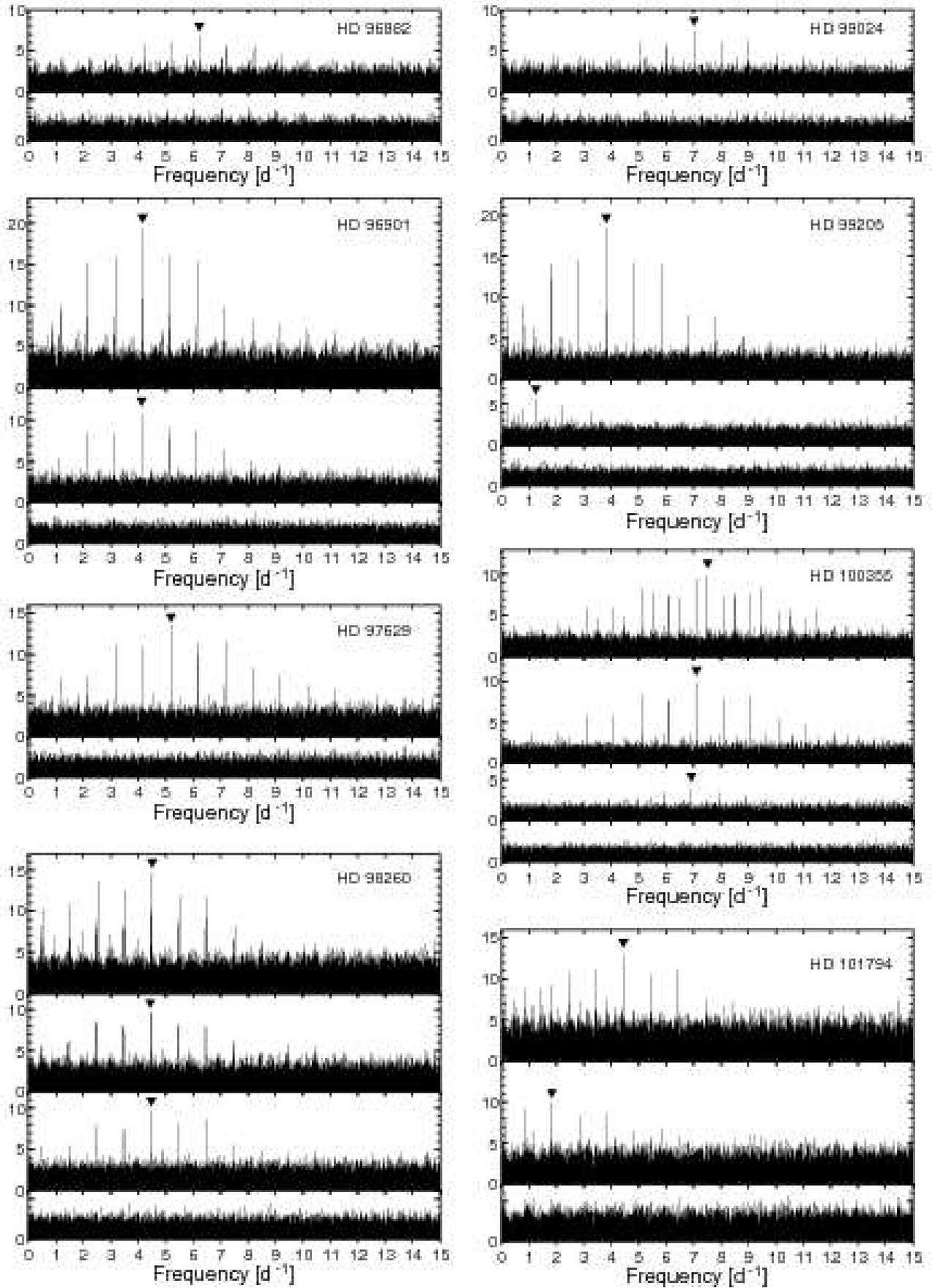}
\caption{The same as in Fig.~\ref{fp-01}, but for HD 96882, 96901, 97629, 98260, 99024, 99205, 100355, and 101794.}
\label{fp-04}
\end{figure*}
}

\onlfig{6}{%
\begin{figure*}
\centering
\includegraphics[width=17cm]{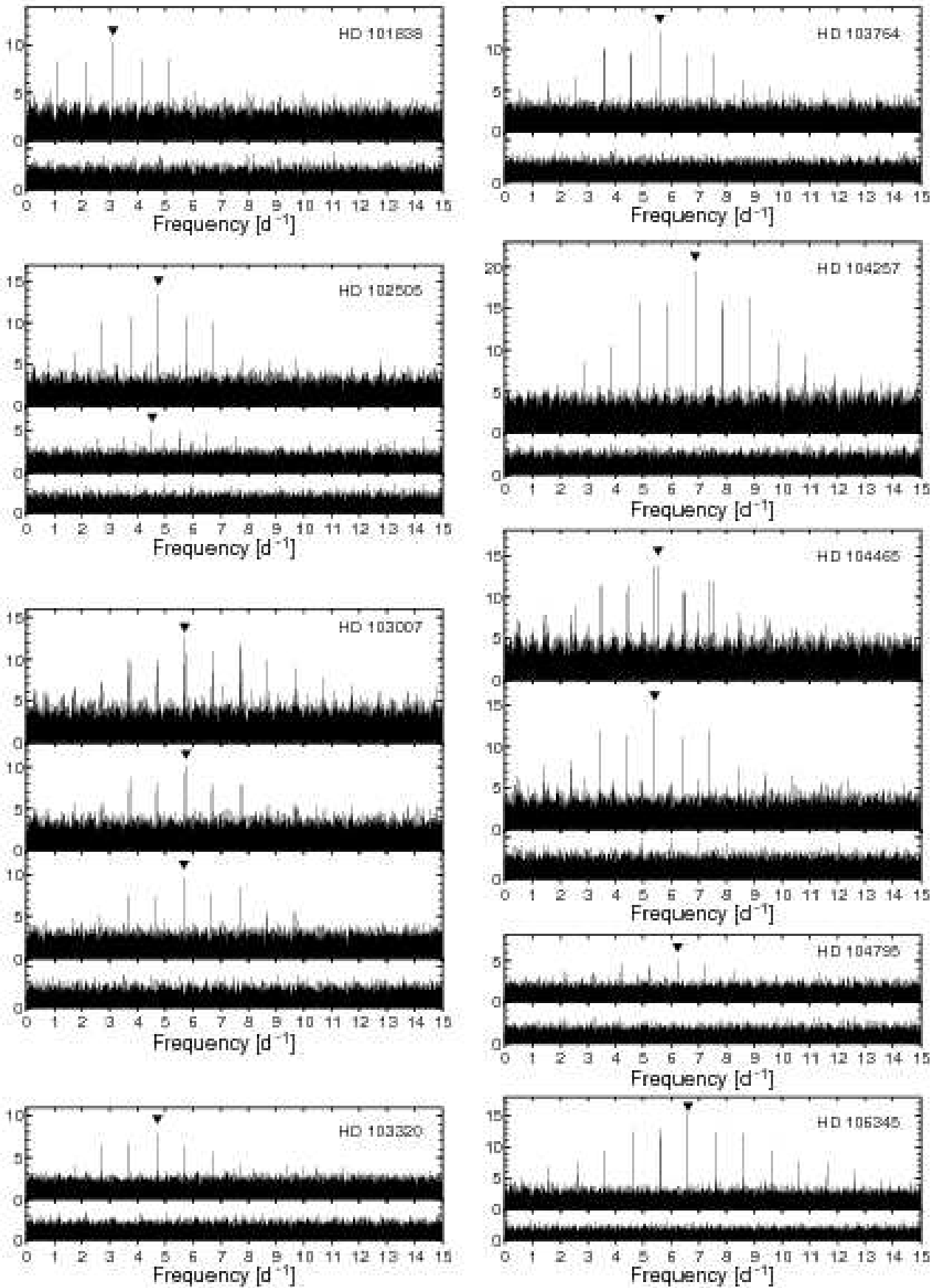}
\caption{The same as in Fig.~\ref{fp-01}, but for HD 101838, 102505, 103007, 103320, 103764, 104257, 104465, 104795, and 106345.}
\label{fp-05}
\end{figure*}
}

\onlfig{7}{%
\begin{figure*}
\centering
\includegraphics[width=17cm]{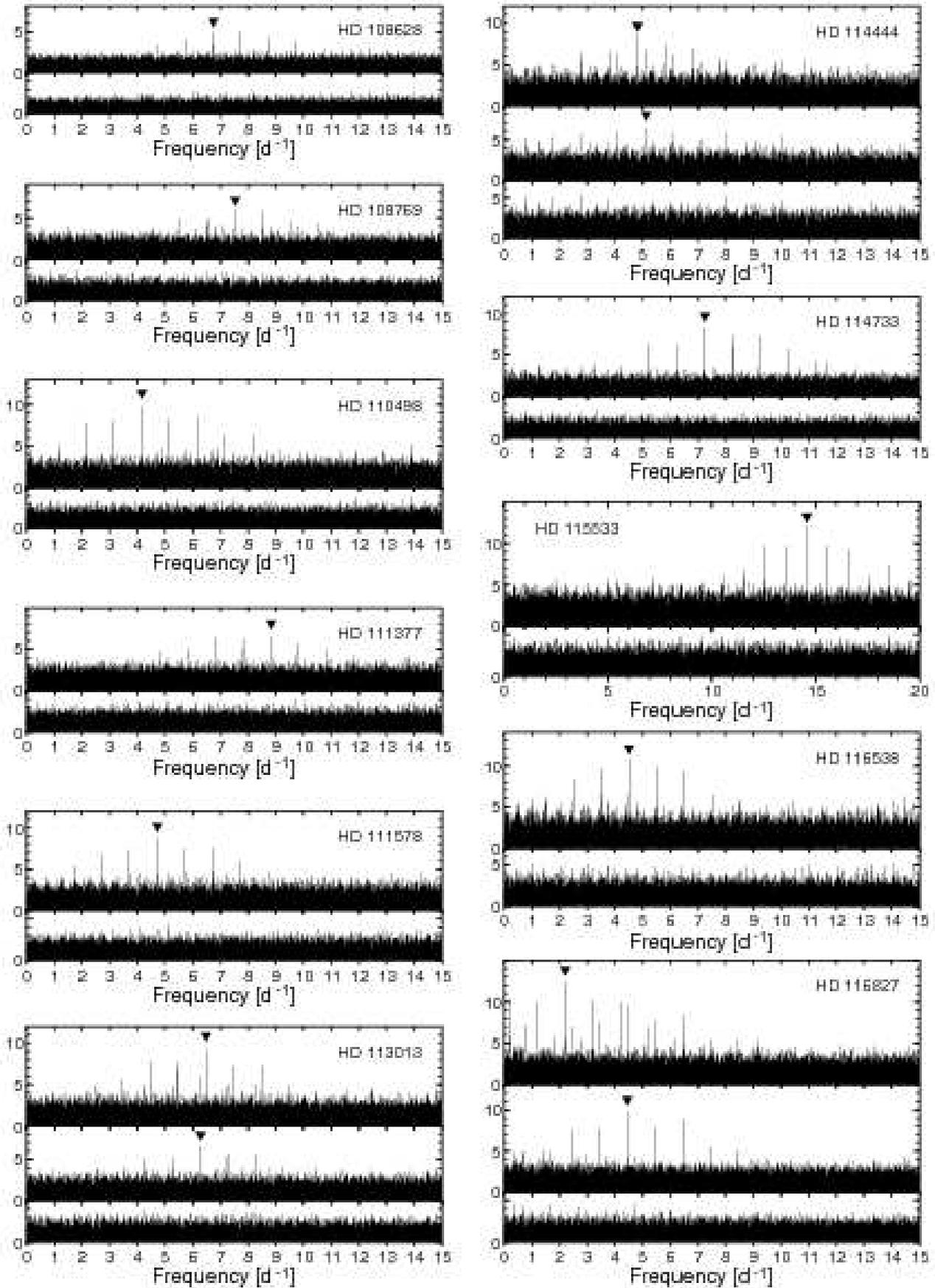}
\caption{The same as in Fig.~\ref{fp-01}, but for HD 108628, 108769, 110498, 111377, 111578, 113013, 114444, 114733, 115533, 116538, and 116827.}
\label{fp-06}7
\end{figure*}
}

\onlfig{8}{%
\begin{figure*}
\centering
\includegraphics[width=17cm]{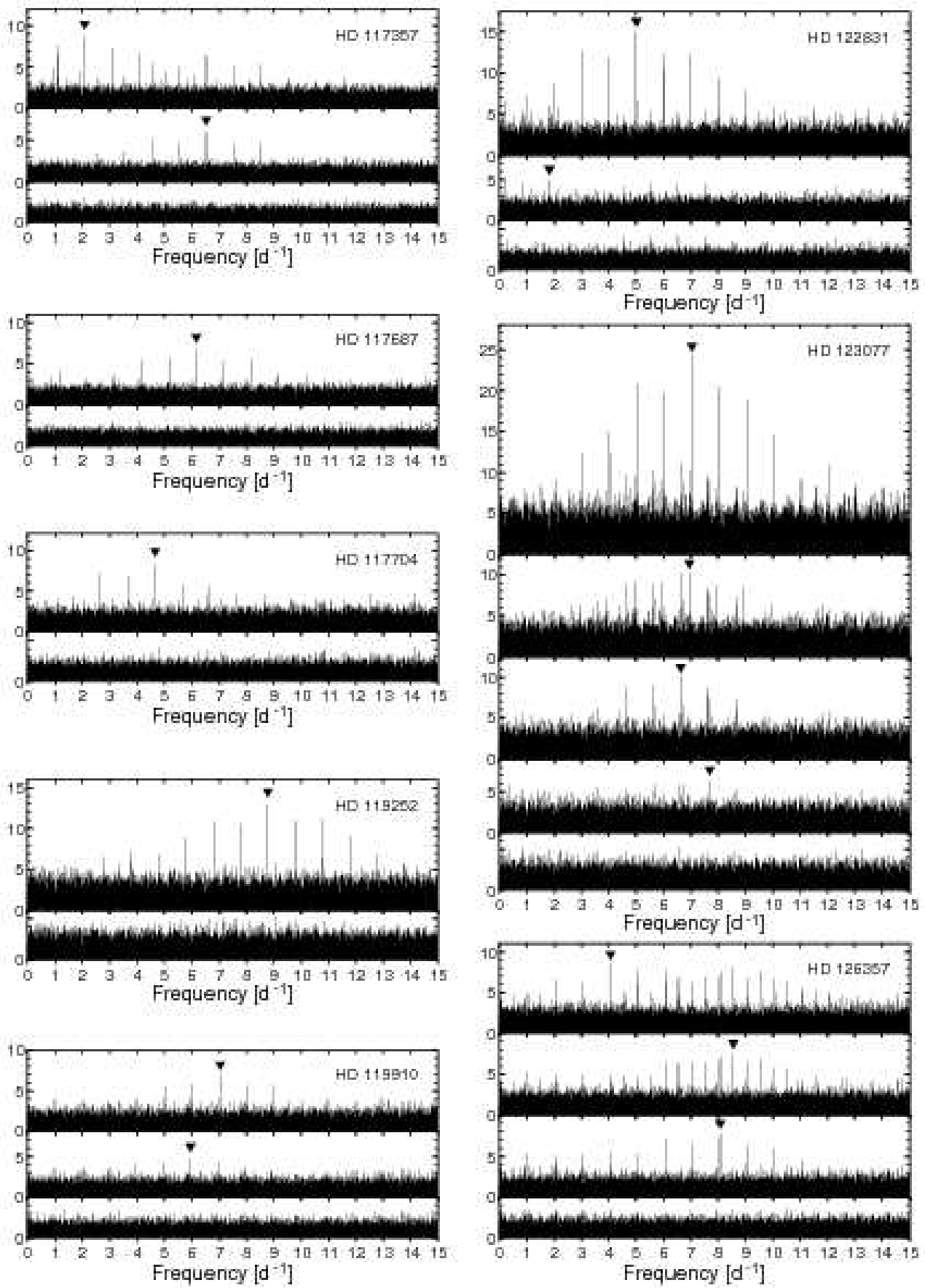}
\caption{The same as in Fig.~\ref{fp-01}, but for HD 117357, 117687, 117704, 119252, 119910, 122831, 123077, and 126357.}
\label{fp-07}
\end{figure*}
}

\onlfig{9}{%
\begin{figure*}
\centering
\includegraphics[width=17cm]{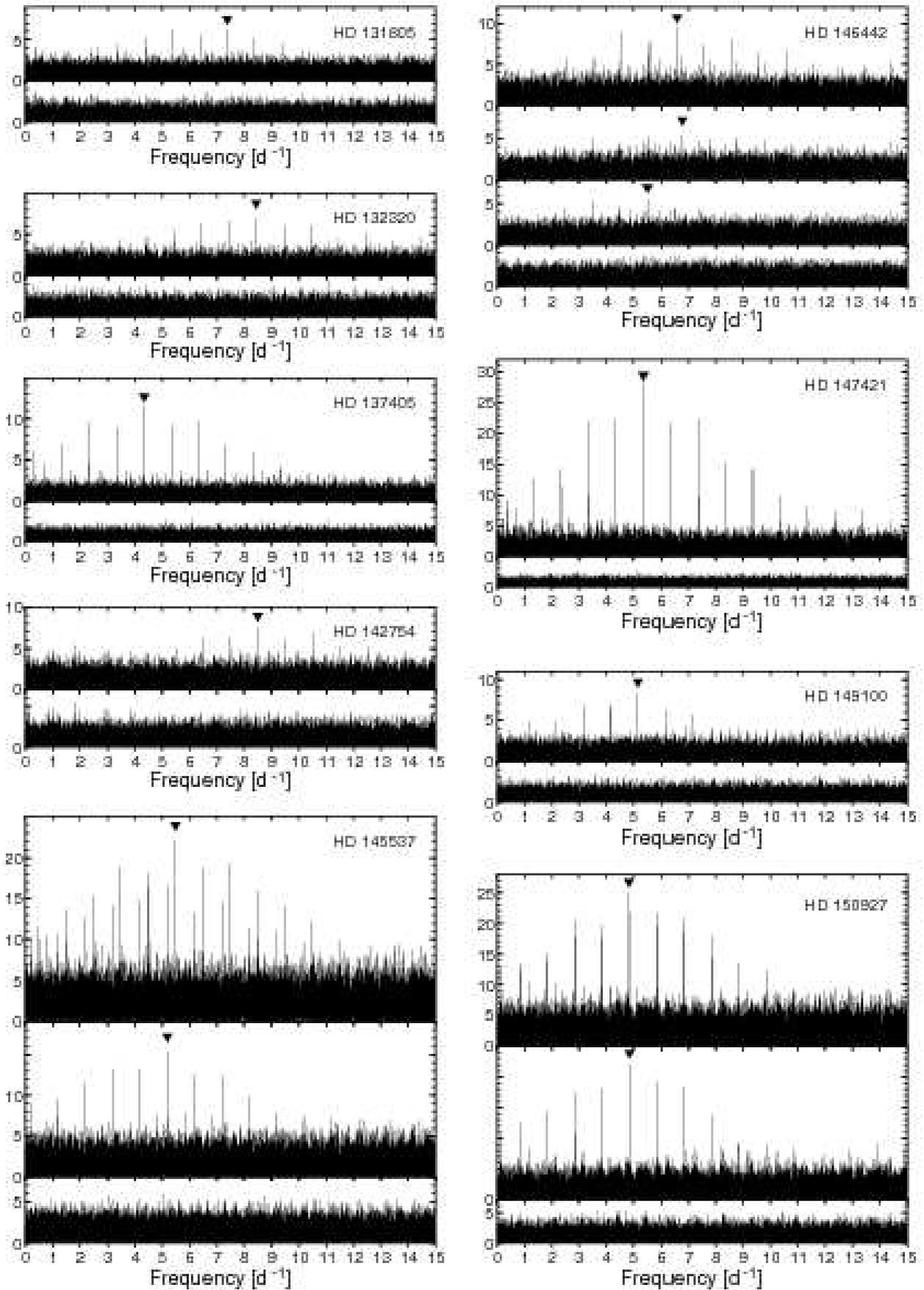}
\caption{The same as in Fig.~\ref{fp-01}, but for HD 131805, 132320, 137405, 142754, 145537, 146442, 147421, 149100, and 150927.}
\label{fp-08}
\end{figure*}
}

\onlfig{10}{%
\begin{figure*}
\centering
\includegraphics[width=17cm]{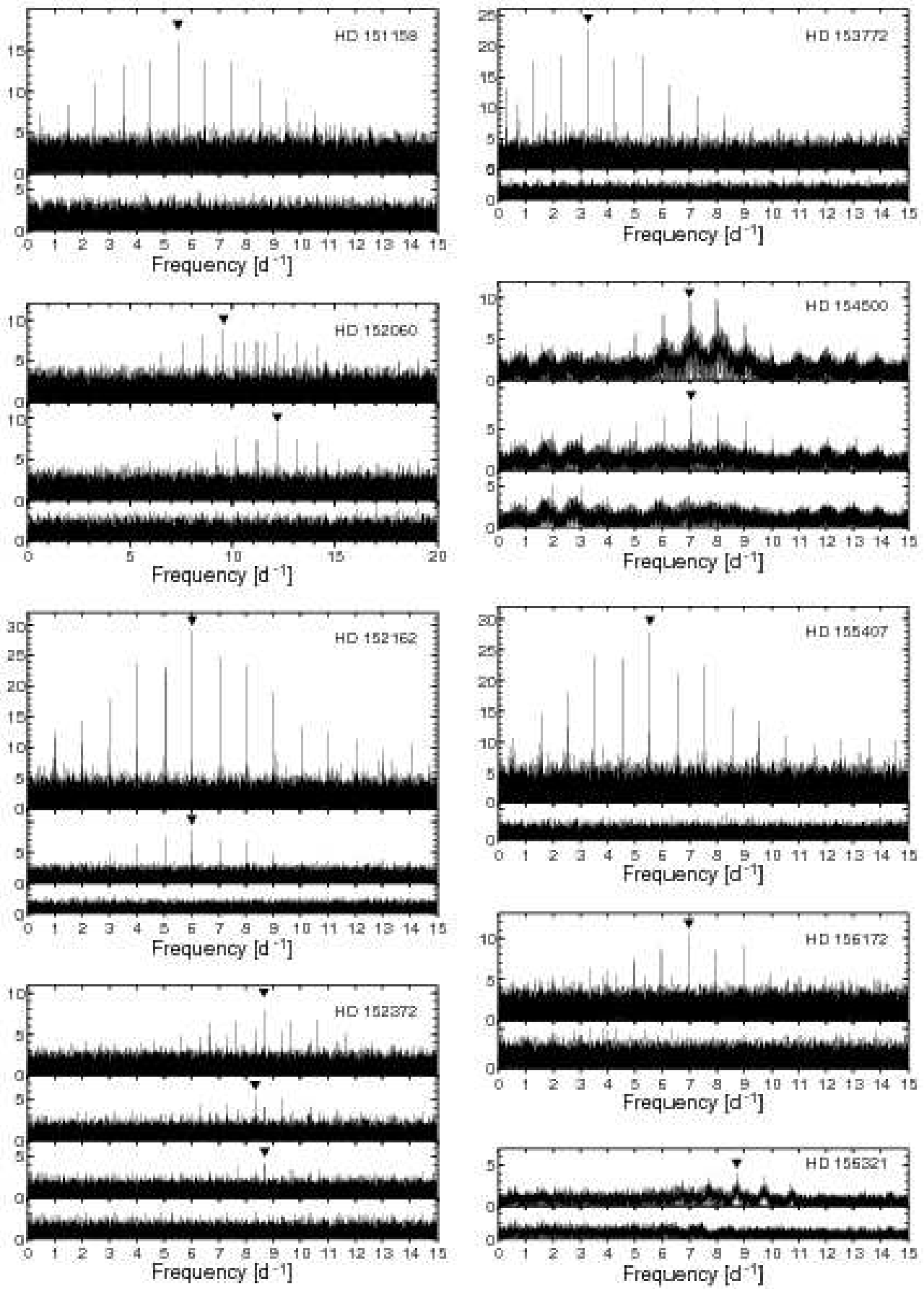}
\caption{The same as in Fig.~\ref{fp-01}, but for HD 151158, 152060, 152162, 152372, 153772, 154500, 155407, 156172, and 156321.}
\label{fp-09}
\end{figure*}
}

\onlfig{11}{%
\begin{figure*}
\centering
\includegraphics[width=17cm]{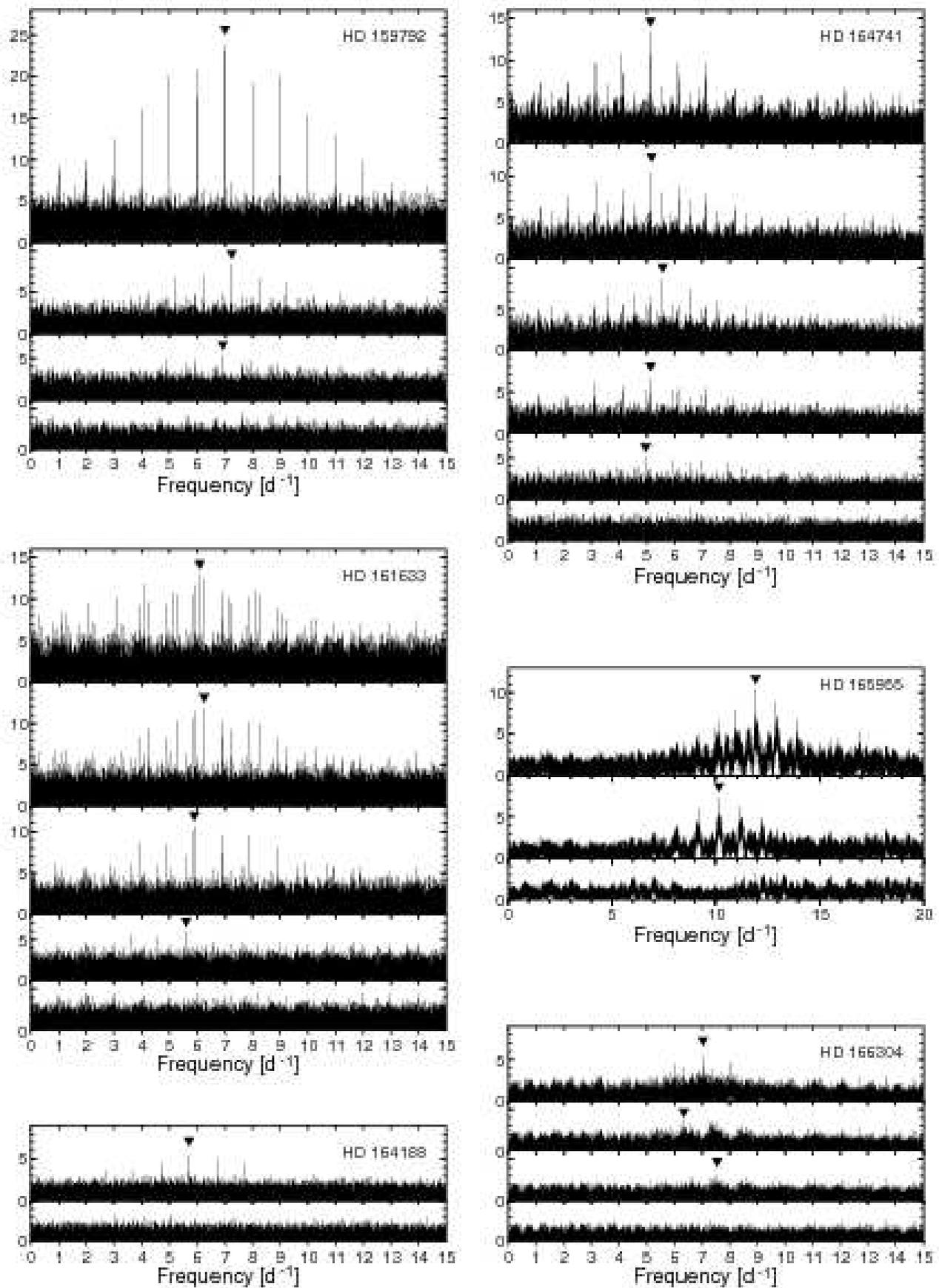}
\caption{The same as in Fig.~\ref{fp-01}, but for HD 159792, 161633, 164188, 164741, 165955, and 166304.}
\label{fp-10}
\end{figure*}
}

\onlfig{12}{%
\begin{figure*}
\centering
\includegraphics[width=17cm]{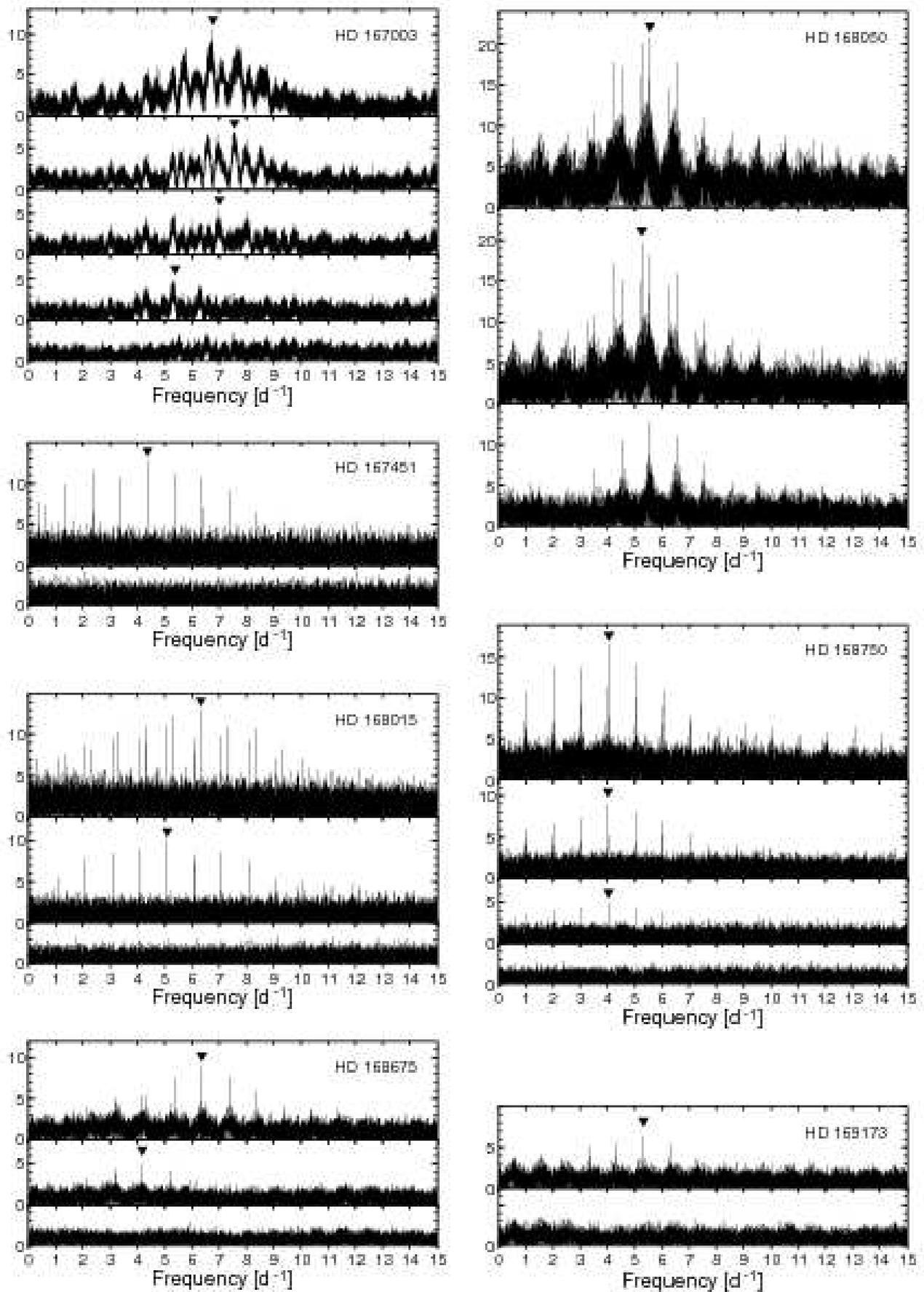}
\caption{The same as in Fig.~\ref{fp-01}, but for HD 167003, 167451, 168015, 168050, 168675, 168750, and 169173.}
\label{fp-11}
\end{figure*}
}

\onlfig{13}{%
\begin{figure*}
\centering
\includegraphics[width=17cm]{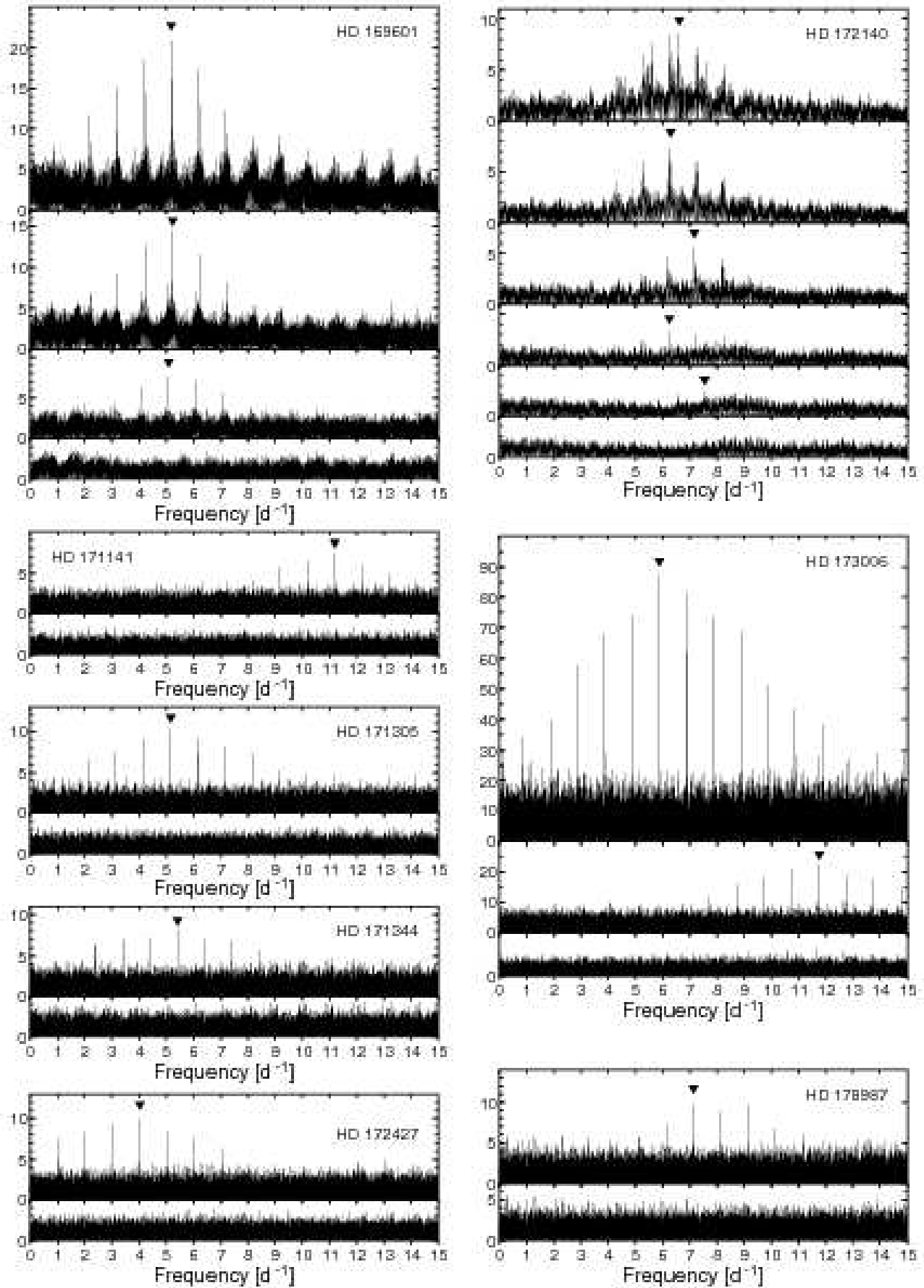}
\caption{The same as in Fig.~\ref{fp-01}, but for HD 169601, 171141, 171305, 171344, 172140, 172427, 173006, and 178987.}
\label{fp-12}
\end{figure*}
}

\onlfig{14}{%
\begin{figure*}
\centering
\includegraphics[width=17cm]{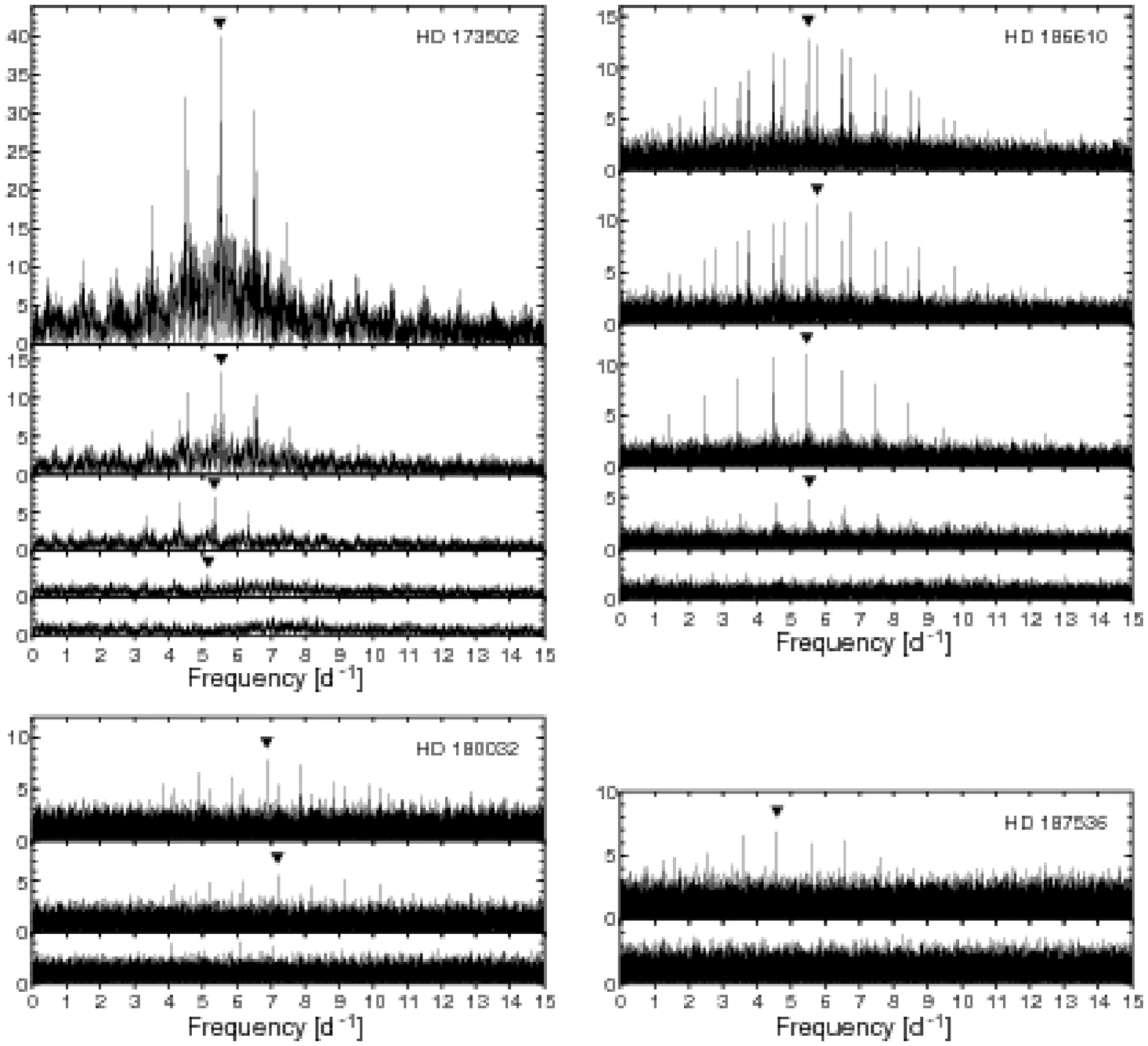}
\caption{The same as in Fig.~\ref{fp-01}, but for HD 173502, 180032, 186610, and 187536.}
\label{fp-13}
\end{figure*}
}

The ASAS-3 photometry we analysed was carried out with a
John\-son $V$ filter. The data cover the time interval between the beginning of the project in 2000 and the end of February, 2006. As already shown in Paper I,
the ASAS-3 data suffer from daily aliases. Consequently, the frequencies of some modes we provide in Tab.~\ref{bc-new-f}, especially of those with the
smallest amplitudes, might be in error by 1 or even 2 cycles per sidereal day. On the other hand, the yearly aliases are relatively low in the ASAS-3 data 
(see Paper I).

\begin{figure}
\includegraphics[width=8.8cm]{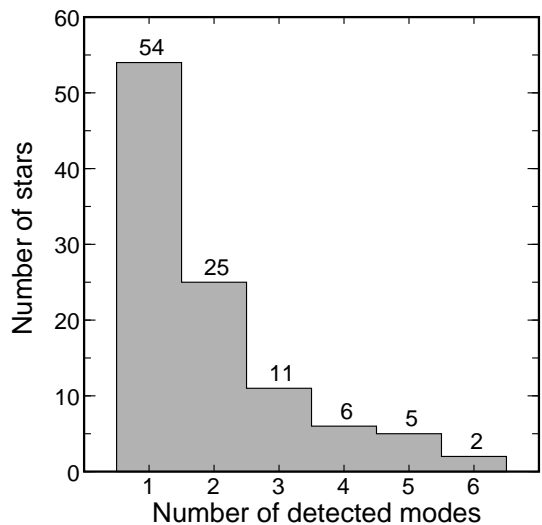}
\caption{The histogram of the number of detected modes for the 103 new $\beta$~Cephei-type stars.}
\label{hist-modes}
\end{figure}

\begin{figure}
\includegraphics[width=8.8cm]{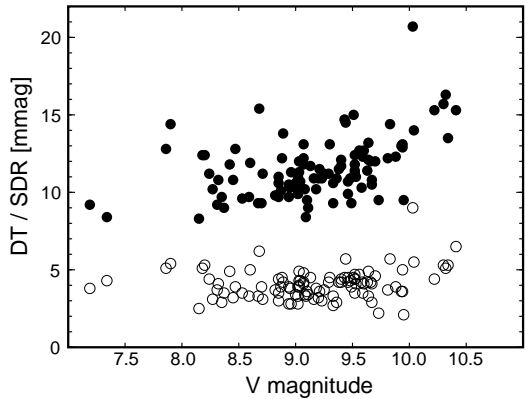}
\caption{Detection threshold, DT (open circles), and standard deviation of the residuals, SDR (filled circles), for the ASAS-3 photometry of the 103 new $\beta$~Cephei stars, plotted against their $V$ magnitude. Detection theshold was defined as corresponding to S/N = 4.}
\label{dett}
\end{figure}

\subsection{Detected modes}
With a detection threshold typical for one-site ground-based data, up to several modes were detected in a single $\beta$~Ce\-phei star.
Currently, the largest number of independent modes, fourteen, was observed
in $\nu$ Eridani as a result of a large photometric campaign \citep{jerz05}. Another $\beta$~Cephei star with large number
of detected modes, namely eleven, is 12 (DD) Lacertae
\citep{hand06}.  A histogram showing the number of detected modes in our sample of the 103 new $\beta$~Cephei stars
is shown in Fig.~\ref{hist-modes}.  In total, 198 independent modes were detected. Up to six modes were found in a single star; there are two
such objects, HD\,74339 and HD\,86214. Some stars with the largest number of detected modes will be discussed in Sect.~3.3.
The detection threshold, DT, defined as four times the average amplitude in the Fourier periodogram of the residuals
in the range 0--40~d$^{-1}$, is shown with open circles in Fig.~\ref{dett} and listed in Tab.~\ref{bc-new-f}.
Typically, it amounts to 3--5 mmag. On the other hand, the standard deviation of the residuals, RSD, which can be regarded as the mean accuracy of a single
measurement, equals to 10--15 mmag, depending on the magnitude (Fig.~\ref{dett}, see also Tab.~\ref{bc-new-f}).
In addition, it can be seen from Fig.~\ref{dett} that all but several
stars have $V$ magnitudes in the range between 8 and 10, i.e., are quite bright. The brightest $\beta$~Cephei-type star we found, at magnitude $V$ = 7.19,
is HD\,149100.

\subsection{Amplitudes and periods}
\begin{figure}
\includegraphics[width=8.8cm]{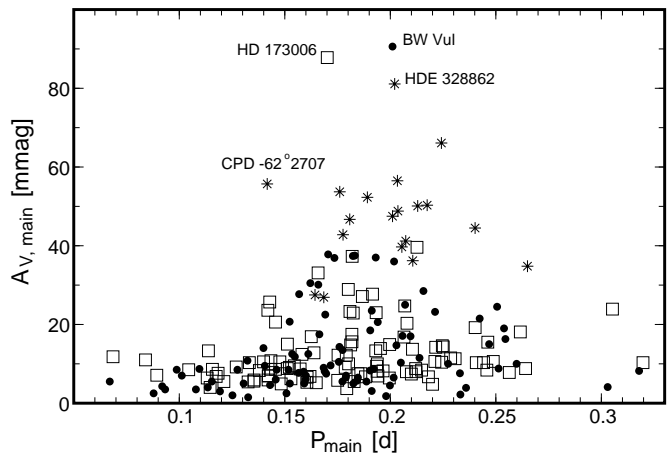}
\caption{$V$-filter semi-amplitude of the main mode, $A_{\rm V, main}$, plotted as a function of the period
of this mode, $P_{\rm main}$. Black dots denote $\beta$~Cephei stars listed by \citet{stha05}; $\alpha$~Vir
and $\omega^1$~Sco are not plotted. Stars discovered by \citet{hand05} and \citet{pigu05} are shown with asterisks, and those listed in Tab.~\ref{bc-new},
as open squares. The four stars discussed in the text are labelled.}
\label{ampl}
\end{figure}

Among stars listed in the review
paper of \citet{stha05}, the well-known $\beta$~Cephei-type BW Vul is exceptional because of its high amplitude.
 As we can see from Fig.~\ref{ampl},
the stars discovered in the ASAS data by \citet{hand05} and \citet{pigu05} fill the gap in amplitudes between BW Vul and the other stars.
One star from that sample, HDE\,328862, has the amplitude similar to that of BW Vul. For another one, CPD\,$-$62$\degr$2707,
a harmonic of the main mode, an indication of non-sinusoidal light curve, was detected.
In our list of newly discovered $\beta$~Cephei stars there is another star with an exceptionally large amplitude, HD\,173006.

\begin{figure*}
\sidecaption
\includegraphics[width=12cm]{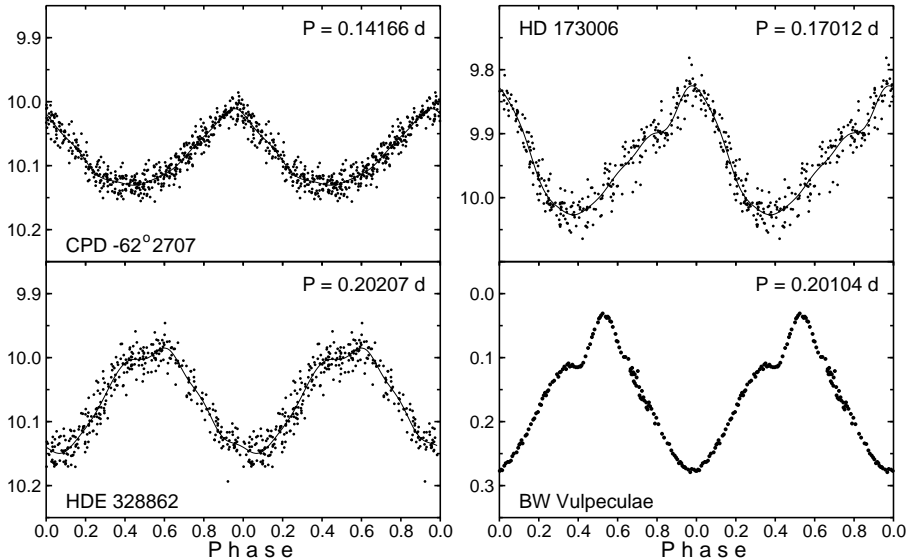}
\caption{Light curves of three large-amplitude $\beta$~Cephei stars discovered in the ASAS-3 data and BW Vulpeculae. Continuous lines
are fits including the main frequency and five lowest harmonics. For the ASAS-3 stars, the ordinate is the
$V$ magnitude, for BW Vul, differential Str\"omgren $b$ magnitude.  Observations of BW~Vul were taken from the 1982 campaign data
\citep{ster86}.}
\label{large-amp}
\end{figure*}

BW Vulpeculae is known for its unusual non-sinusoidal light curve, with a standstill on the raising branch (Fig.~\ref{large-amp}). It is interesting to note that a standstill was also found in the low-amplitude mode of 15 (EY) CMa \citep{shob06}.
In Fig.~\ref{large-amp}, we plot phase diagrams for the main modes of BW Vul and the three large-amplitude
stars mentioned above.  It can be seen that HD\,173006 and HDE\,328862 also exhibit a standstill
although it is not so clearly visible as for BW Vul due to the larger scatter of datapoints. The light curve for BW Vul shown in Fig.~\ref{large-amp}
was obtained in Str\"omgren $b$ filter, but it is very similar in shape to that in $V$ \citep[see, e.g.,][]{ster93}.

As can be seen in Fig.~\ref{ampl}, the periods of modes detected in the new $\beta$~Cephei stars match very well the range covered by
already known stars. It seems that the largest amplitudes occur in stars with intermediate periods; for
the shortest and the longest periods amplitudes are rather small. This result qualitatively agrees with the recent non-linear calculations of \citet{smmo07}
for radial modes in $\beta$~Cephei stars.

There were ten known $\beta$~Cephei stars having modes with periods shorter than 0.1~d, of which six are members of the open cluster
NGC\,6231 \citep[see, e.g.,][]{bala95}. We detect four next ones, HD\,115313, 152060, 165955, and 171141.  The shortest
period, 0.0686~d, was found in HD\,115313.   It is interesting to note that
HD\,152060 = Braes 122 = NGC\,6231-308 is located 28$\arcmin$ off the centre of NGC\,6231. It is thus possible that this star is a member of the
cluster or of the Sco OB1 association surrounding NGC\,6231.

On the other side of the period range we knew two stars with periods slightly longer than 0.3~days, V986 Oph  with period equal to 0.303~d \citep{cuyp89}
and Oo\,2299 in the open cluster $\chi$ Persei with period of 0.31788~d, \citep{krpi97}.
We found two stars with periods longer than 0.3~d (see, however, Sect.~3.4): HD\,101838 ($P$ = 0.31973~d, a member
of the open cluster Stock 14 and a component of an eclipsing binary, see Sect.~3.7) and HD\,153772 = vdBH~81a ($P$ = 0.30527~d), in the core of
the R association vdBH 81 \citep{vdbh75,herb75}.

As we already explained in the Introduction, we expected to detect $\beta$~Cephei stars with smaller amplitudes than those
found earlier in the ASAS data by \citet{hand05} and \citet{pigu05}. Fig.~\ref{ampl} shows that this is indeed the case. 

\subsection{Multiplicity and splittings}
\begin{figure}
\includegraphics[width=8.8cm]{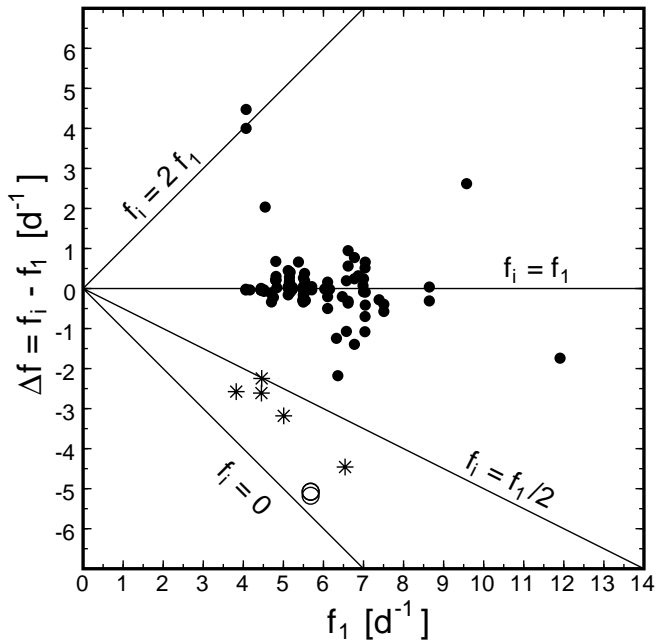}
\caption{The difference, $\Delta f$ = $f_{\rm i} - f_1$, where $f_{\rm i}$ is the frequency of a given mode and $f_1$, the frequency of the main mode,
for 49 multiperiodic $\beta$~Cephei stars discovered in this paper. Low-frequency modes (see Sect.~3.4) are shown as asterisks. In addition,
long-period modes for HD\,133823 (Paper I) are shown with open circles.}
\label{fdf}
\end{figure}

As mentioned in Sect.~3.1 (see also Fig.~\ref{hist-modes}), about half of the new $\beta$~Cephei stars are multiperiodic. In order to show how the
modes are spread over the frequencies, in Fig.~\ref{fdf} we plot the differences, $\Delta f$, between frequency of a given mode, $f_{\rm i}$, and the
frequency of the main mode, $f_1$, as a function of $f_1$. By the main mode we mean the mode with the largest amplitude and period shorter than
0.32~d. We see from Fig.~\ref{fdf} that $\Delta f$ is usually small in comparison with $f_1$ and the value of $|\Delta f|/f_1$ rarely exceeds
0.2;  in most cases it is much smaller. There are, however, a few exceptions. First, for HD\,126357, the frequencies of two modes, $f_2$ = 8.54271~d$^{-1}$
and $f_3$ = 8.07010~d$^{-1}$, are about twice the frequency of the main mode ($f_1$ = 4.06927~d$^{-1}$). This situates these two modes close to
the $f_{\rm i}$ = 2$f_1$ line in Fig.~\ref{fdf}. Next, we have five stars with modes that
are placed below $f_{\rm i}$ = $f_1$/2 line (asterisks in Fig.~\ref{fdf}). We will discuss these stars in Sect.~3.4.

Rotational splittings are observed in some $\beta$~Cephei stars. Good examples are $\nu$~Eri \citep{jerz05}, 
V836 Cen \citep{aert04}, and $\theta$~Oph \citep{hasm05}.
The presence of two modes closely spaced in frequency does not prove that we deal with rotational splitting although 
this possibility
has to be considered in such a case. When three or more equally (or almost equally) spaced modes are observed,
the rotational splitting is an obvious conclusion. However, it can happen accidentally that three modes of different $\ell$ 
form an equidistant triplet, as found for 12 (DD) Lac \citep{hand06}.

\begin{figure*}
\includegraphics[width=18cm]{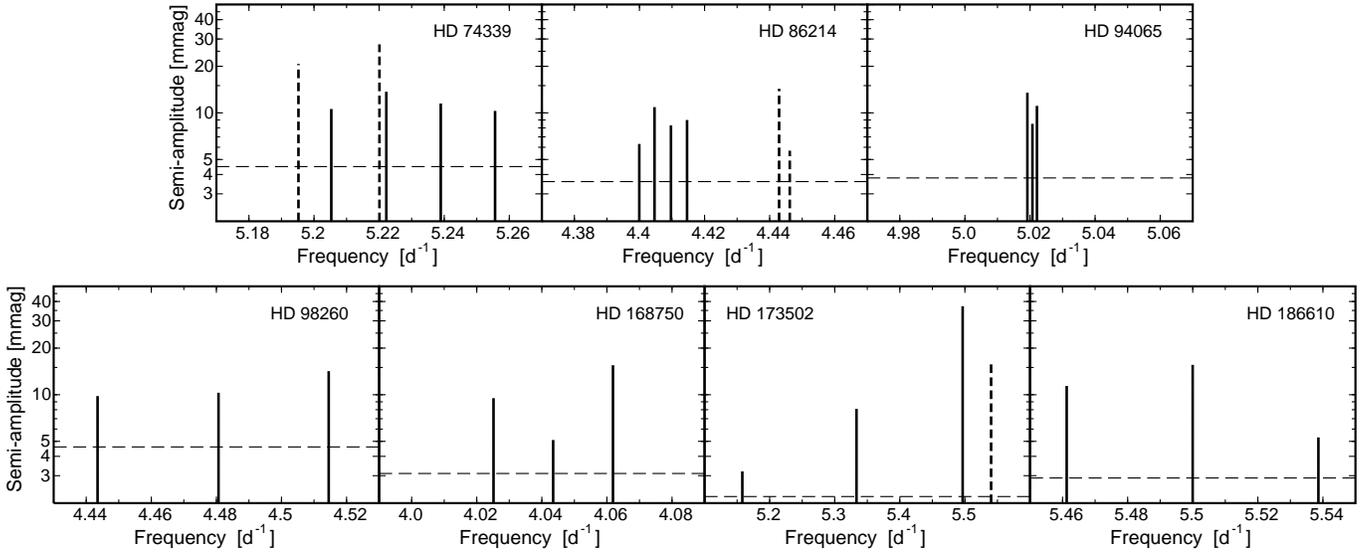}
\caption{Schematic frequency spectra in the vicinity of the possible rotationally split modes of seven new $\beta$~Cephei stars. For all stars but HD\,173502
the frequency range shown is equal to 0.1~d$^{-1}$. Modes which are not components of the suggested multiplets are plotted with dashed lines.
See also Tab.~\ref{splittings}.}
\label{multi}
\end{figure*}

The rotationally split modes are potentially very useful for mode identification and asteroseismology. 
We therefore checked the frequencies of the detected modes in the new $\beta$~Cephei stars
for possible rotational splittings. The results are summarized in Tab.~\ref{splittings} and shown in Fig.~\ref{multi}.
We list only stars in which inequality of separations is smaller than 20\%. The most interesting are the first
two stars in Tab.~\ref{multi}, HD\,74339 and HD\,86214, those that in our sample have the largest number of detected modes. In both
stars equidistant quadruplets are observed. It is quite likely that modes listed in Tab.~\ref{splittings} are parts 
of a rotationally split $\ell$ = 1 or 2 modes. This, however, needs to be verified by mode identification, as 
was done for 12 Lac.

\begin{table}
\caption{Possible rotational splittings for seven stars from our sample.
Designation of frequencies is the same as in Tab.~\ref{bc-new-f}.}
\label{splittings}
\centering
\begin{tabular}{rcll}
\hline\hline
Star & $\Delta f$ & Splitting [d$^{-1}$] & Note\\
\hline
HD\,74339 & $f_3 - f_5$ & 0.01690(4) & quadruplet\\
& $f_4 - f_3$ & 0.01673(4) &\\
& $f_6 - f_4$ & 0.01669(4) &\\
HD\,86214 & $f_2 - f_5$ & 0.00468(5) & quadruplet\\
& $f_3 - f_2$ & 0.00505(4) &\\
& $f_4 - f_3$ & 0.00495(4) &\\
HD\,94065 & $f_5 - f_3$ & 0.00155(4) & triplet\\
& $f_2 - f_5$ & 0.00139(4)&\\
HD\,98260 & $f_3 - f_2$ & 0.03716(4) & triplet\\
& $f_1 - f_3$ & 0.03386(4) &\\
HD\,168750 & $f_3 - f_2$ & 0.01833(2) & triplet\\
& $f_1 - f_3$ & 0.01838(4) &\\
HD\,173502 & $f_3 - f_4$ & 0.17543(9) & triplet\\
& $f_1 - f_3$ & 0.16308(3) &\\
HD\,186610 & $f_1 - f_3$ & 0.03871(4) & triplet\\
& $f_4 - f_1$ & 0.03863(5) &\\
\hline
\end{tabular}
\end{table}

\subsection{Stars with long periods}
The longest periods observed in Galactic $\beta$~Cephei stars rarely exceed 0.3~d (see Fig.~\ref{ampl}) and until recent years there was no clear evidence for
the presence of low-frequency $g$ modes in these stars.  From the theoretical point of view, however, some overlap of the instability strips
of $\beta$~Cephei and slowly pulsating B (SPB) stars \citep[e.g.,][]{pamy99} indicated that some hybrid objects might be found.
The recent campaigns on two well-known $\beta$~Cephei stars, $\nu$~Eri \citep{jerz05} and 12 (DD) Lac \citep{hand06},
led to the discovery of low-frequency mode(s) in these stars.

On the other hand, \citet{kola04, kola06} announced the discovery of a large number of short-period variable stars
in the Magellanic Clouds showing long-period mode(s) (periods in the range of 0.4--1 d) in addition to mode(s) typical for $\beta$~Cephei stars.
The origin of these long-period modes still needs to be explained. We will come back to the discussion of this problem in Sect.~4.

In the ASAS-3 data we have found five stars having periods considerably longer than 0.3~d (Fig.~\ref{fdf}), namely, in the range between 0.45 and 0.80~d.
These stars are listed in Tab.~\ref{long}. The most interesting of the five stars listed in Tab.~\ref{long} is HD\,101794 = V916\,Cen = Stock14-13. The star is
a member of the loose open cluster Stock\,14 = Lod\'en 462/465 \citep{lode73}, although \citet{hump78} and \citet{kage94} include it to the association Cru OB1.
The star was classified as B1\,IVne by \citet{garr77}.  Its Be status was confirmed by measurements of the $\beta$ index \citep{movo75,klne77,joha81,kalt03}
and H$\alpha$ photometry \citep{mcgi05}.

\begin{table}
\caption{Stars with periods longer than 0.4~d.}
\label{long}
\centering
\begin{tabular}{cccl}
\hline\hline
& Short & Long &\\
Star &  period [d] & period [d] & Note\\
\hline
HD\,99205 & 0.26161 & 0.80203 & \\
HD\,101794 & 0.22447 & 0.54362 & V916 Cen, Be\\
HD\,116827 & 0.22429 & 0.45217 & Be\\
HD\,117357 & 0.15291 & 0.48052 & Be\\
HD\,122831 & 0.19968 & 0.54730 &\\
\hline
\end{tabular}
\end{table}

HD\,101794 was discovered as variable by Hipparcos and was subsequently named V916\,Cen. The star was assigned the variability type of $\gamma$~Cas,
which labels B-type stars showing erratic long-term changes with a range up to 0.5~mag.  Suprisingly, HD\,101794 is also an eclipsing star (see Sect.~3.7).
There are at least two other periodic terms superimposed
on the variability due to eclipses (Tab.~\ref{long}).  One of them has a period in the $\beta$~Cephei domain ($P$ = 0.22447~d) and this is the reason for
including the star in the list of $\beta$~Cephei stars.   Another periodicity, with a period
of about 0.54362~d, can be attributed to a $\lambda$~Eridani-type variability.

Another star from Tab.~\ref{long}, HD\,116827, is also known to be a Be star \citep{macc81}.
\citet{houk1} classify the star as B2/5 (Vn) and
note that the H$\beta$ line may be partially filled in, a clear indication of a weak emission.

The third new $\beta$~Cephei star with both short and long-period variability, is HD\,117357. It shows even stronger emission
than HD\,101794 in hydrogen Balmer lines \citep{schm64,garr77,wiga98,yudi01}.
The emission in H$\beta$ is confirmed by two measurements of the $\beta$ index: 2.522 \citep{deut76}
and 2.502 \citep{klne77}. 

There is no recorded evidence for emission in hydrogen lines for the two remaining stars 
in Tab.~\ref{long}, HD\,99205 and HD\,122831. 
Since three of five stars in Tab.~\ref{long} show Balmer-line emission and only two of the remaining 98 $\beta$~Cephei stars are known as
Be stars, it is hard to believe that the occurence of long periods and Balmer-line emission remains unrelated.
We will come back to the discussion of the origin of long periods in these stars in Sect.~4.

\subsection{Period and amplitude changes}
\begin{figure}
\includegraphics[width=8.8cm]{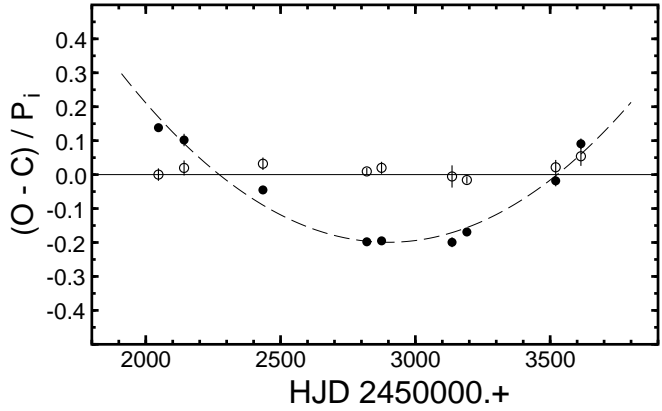}
\caption{The O$-$C diagram for both modes detected in HD\,168050. The dashed line represents constant rate of period change
equal to 104~s\,century$^{-1}$.}
\label{168050oc}
\end{figure}

Period and amplitude changes are observed in some $\beta$~Cephei stars \citep[][Paper I]{jepi98};
they usually take place on a time scale of years or decades. Because ASAS-3 data cover five years, we 
may expect detecting period changes only if they are very fast. Indeed, only one star, HD\,168050, shows 
pronounced period changes. The photometry of HD\,168050 reveals not only two modes attributable to the
$\beta$~Cephei-type variability (Tab.~\ref{bc-new-f}), but also eclipses (see Sect.~3.7), indicating that
the pulsating star is a component of an eclipsing binary. For one of the detected pulsation modes, the period is stable,
while for the other the period increases rapidly.
We show the O$-$C diagram for both modes in Fig.~\ref{168050oc}. Assuming constant rate of period change for $P_2$, we get
$dP_2/dt$ = 104 $\pm$ 8 s century$^{-1}$. This is the fastest change of period ever detected in a $\beta$~Cephei-type star.
Times of maximum light for both modes of HD\,168050 are given in Tab.~\ref{168050tmax}.

\begin{table}
\caption{Time of maximum light for both modes detected in HD\,168050.}
\label{168050tmax}
\centering
\begin{tabular}{cc}
\hline\hline
$T_{\rm max,1}$ &$T_{\rm max,2}$\\
\multicolumn{2}{c}{[HJD\,2450000.0+]}\\
\hline
2047.1099(20) & 2047.0437(33)\\
2142.0830(33) & 2142.0758(40)\\
2434.5649(21) & 2434.5908(33)\\
2819.8627(12) & 2819.8428(18)\\
2874.6522(18) & 2874.5004(32)\\
3135.9806(26) & 3135.9672(59)\\
3190.9553(18) & 3191.0019(19)\\
3520.4374(29) & 3520.4664(39)\\
3613.9949(28) & 3613.9773(51)\\
\hline
\end{tabular}
\end{table}

As far as the long-term amplitude changes are concerned, we found no clear evidence for them among the newly discovered $\beta$~Cephei stars.

\subsection{O-type stars}
Although B0 is the earliest spectral type of known $\beta$~Cephei stars, from the theoretical point of view the instability strip 
of $\beta$~Cephei stars extends to O-type stars \citep[see, e.g.,][]{pamy99}. The searches for $\beta$~Cephei-type
pulsations in O-type stars, both in the Galactic field \citep{balo92} and Cygnus OB2 association \citep{piko98},
were not conclusive, however.  On the other hand, line-profile variations were observed in some fast-rotating late O-type stars 
such as $\zeta$~Oph \citep{vope83, kamb90}. They were usually attributed to pulsations in modes with relatively high $\ell$, undetectable
in photometry.   A recent MOST satellite photometry of $\zeta$~Oph \citep{walk05} clearly revealed $\beta$~Cephei-type pulsations.
It is therefore reasonable to regard $\zeta$~Oph and several other late O-type stars showing non-radial pulsations
in line profiles \citep{full96,badz99} as $\beta$~Cephei stars.

Among the new $\beta$~Cephei stars, there are at least two O-type stars. The first one is HD\,91651 = HIP\,51681, a member
of Car OB1 association. Its published MK spectral types are the following:
O9\,Vp? \citep{morg55}, O9\,Vn: \citep{feas57}, O9\,V:n \citep{walb73}, O8/9\,V \citep{houk1},
O9\,Vnp \citep{garr77}, and O9.5\,III \citep{math88}. The star is fast-rotating, the projected radial velocity amounting
280--300~km\,s$^{-1}$ \citep{penn96,howa97,penn04}. The star is known to have strong nitrogen
lines for its spectral type \citep{garr77,schi85,hopr89}. The variability of radial velocity of HD\,91651
was indicated by \citet{leva88} and \citet{lehn03}. \citet{leva88} suspected even that the star is a double-lined 
spectroscopic binary with an orbital period of 9.275 d. It is interesting to note that from the study of line profiles in this star, \citet{penn96}
already suggested the possibility of non-radial pulsations.

The other O-type star is HD\,156172 with two published spectral types: O9.5/B0.5\,Ia/Iab \citep{houk2} and O9\,Vnne \citep{garr77}. The
star is located in the region of the HI bubble surrounding O5\,Vn((f)) star HD\,155913, and could belong to an unknown association 
\citep{cabe98}.

Two other stars are classified as O9/B0\,V and O9.5/B0\,V by \citet{houk1}, namely,  HD\,95568 and HD\,117357.
For HD\,95568 this is the only MK spectral type available, for HD\,117357 there are several other classifications indicating 
that it is rather a fast-rotating B0 star.

Although we cannot exclude that variability we detected in O-type stars is due to the pulsations of their,
presently unknown, early B-type companions, this seems to be rather unlikely.

\subsection{Components of detached eclipsing binaries}
If a $\beta$~Cephei star is a component of an eclipsing system, one has an opportunity of deriving its accurate mass and radius.
Of the 112 previously known $\beta$~Cephei stars, there were four components of eclipsing systems \citep[see][]{pigu06}:
16 (EN) Lac \citep{jerz78,pije88}, V381\,Car in NGC\,3293 \citep{enba86,jest92,frey05}, $\eta$~Ori \citep{kust16,wala88,deme96}, 
and $\lambda$~Sco \citep{shlo72,uytt04,brbu06}.

We add four stars to this list. Surprisingly, two are members of a loose open cluster Stock 14 \citep{movo75,fimi83,pefi88}:
HD\,101794 and HD\,101838.  The other two are HD\,167003 and HD\,168050. It is interesting to note that at least in two
stars, HD\,101794 and HD\,168050 the secondary eclipses are deep enough to expect that spectroscopically they will be double-lined
systems. If this were the case, the follow-up studies should allow deriving the masses and radii of the components.

\subsubsection{HD\,101794 = ALS\,2460 = HIP\,57106 = V916~Cen = Stock 14-13}
\begin{figure}
\includegraphics[width=83mm]{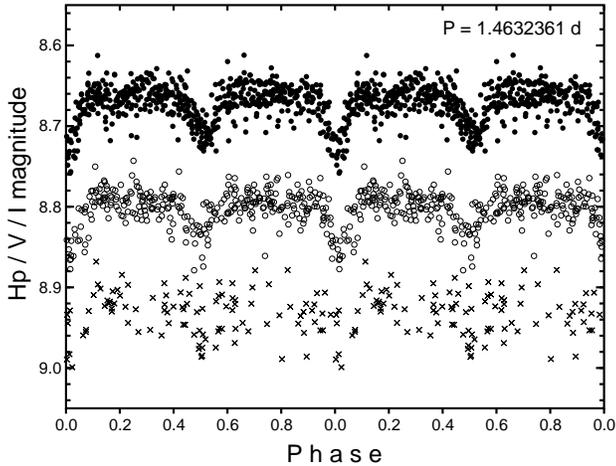}
 \caption{The eclipsing light curve of V916\,Cen folded with the orbital period of 1.4632361~d for ASAS-2 $I$-filter data (dots), ASAS-3
$V$-filter data (open circles) and Hipparcos $H_{\rm p}$ data (crosses). The long-term changes and the contribution
from pulsation were removed.}
\label{fig1}
\end{figure}

As mentioned in Sect.~3.4, photometric variability of HD\,101794, a Be star and a member of Stock\,14, was discovered by the Hipparcos satellite.
Its eclipsing nature was revealed by \citet{pojm00} with the $I$-filter ASAS-2 data. The orbital period
amounted to 1.4632~d. As already explained in Sect.~3.4, the star shows also low-amplitude periodic variations.
The eclipsing light curve of HD\,101794 is shown in Fig.~\ref{fig1}. We see that after removing the contribution from the long-term and
sinusoidal variations, the eclipses are clearly seen both in the 1997--1999 ASAS-2 data and the ASAS-3 data. 
Although barely, they can be even seen in the Hipparcos data.
The ephemerides for eclipses of V916~Cen and the other three stars are given in Tab.~\ref{tab-eph}.

\begin{table*}
 \caption{Data for the four new $\beta$~Cephei stars in eclipsing systems.}
  \begin{tabular}{@{}clll@{}}
 \hline\hline
HD & \multicolumn{1}{c}{$P_{\rm orb}$ [d]} &  \multicolumn{1}{c}{HJD of $T_{\rm min,0}$}  & Remarks \\
 \hline
101794 &  ~~1.4632361(21) & 2450561.9213(15) &  V916\,Cen, Be star, member of Stock 14\\
101838 &  ~~5.41166(7) & 2452014.633(13) & member of Stock 14\\
167003 &  10.79824(11) & 2452692.824(11) &  V4386\,Sgr\\
168050 &  ~~5.02335(6) & 2452096.224(10) & \\
\hline
\end{tabular}
\label{tab-eph}
\end{table*}

\begin{figure}
\includegraphics[width=83mm]{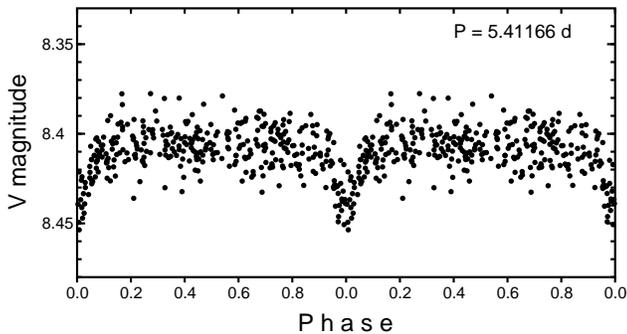}
 \caption{The ASAS-3 $V$-filter light-curve of HD\,101838, freed from the contribution of pulsation and folded
with the orbital period of 5.41166~d.}
\label{fig2}
\end{figure}

\subsubsection{HD\,101838 = ALS\,2463 = Stock 14-14}
HD\,101838 is also a member of the same cluster, Stock~14. Like in V916\,Cen, presumably the primary component is
a $\beta$~Cephei-type star. The pulsation period revealed from the ASAS-3 photometry is quite long, amounting to
0.31973~d. The star is classified as B1\,III \citep{feas61,walk63}, B0.5/1\,III \citep{houk1}, 
B1\,II-III \citep{garr77}, or B0\,III \citep{fimi83}.

The eclipsing light-curve of HD\,101838 is shown in Fig.~\ref{fig2}. When folded with the period of 5.41166~d, it shows no sign of a
secondary minimum. However, we cannot exclude the possibility that the true orbital period is twice as long, i.e., it is equal to about 10.82~d.
If this were the case, the system would consist of two similar components.

\begin{figure}
\includegraphics[width=83mm]{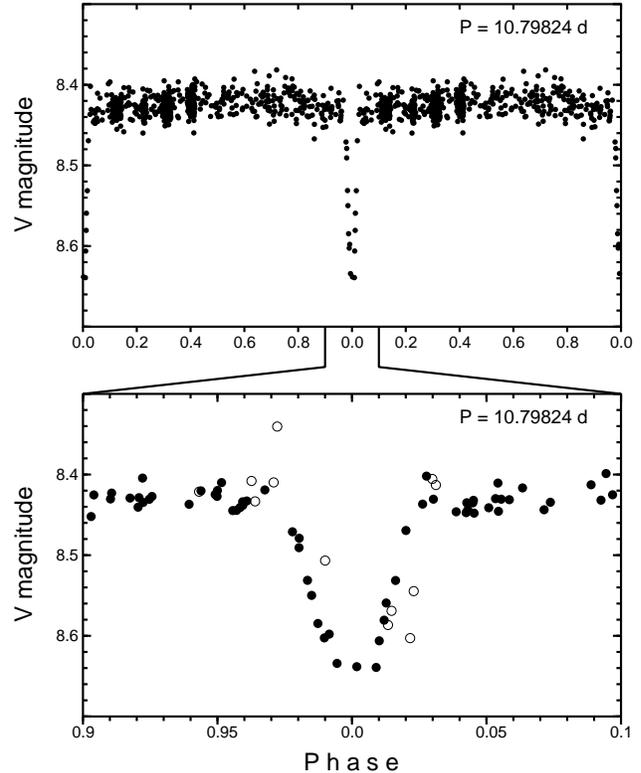}
 \caption{The ASAS-3 $V$-filter light curve of HD\,167003 (V4386\,Sgr) freed from the contribution of pulsations 
and folded
with the orbital period of 10.79824~d. The upper figure shows the whole light curve, the lower, only phases close to
the primary minimum. In addition, Hipparcos data (not corrected for the contribution from pulsations) are plotted in the lower figure as open circles. }
\label{fig3}
\end{figure}

\subsubsection{HD\,167003 = ALS\,4801 = HIP\,89404 = V4386\,Sgr}

The variability of HD\,167003 was detected by the Hipparcos satellite. Since five observational points obtained within half a day
deviated by about 0.2~mag from the remaining observations, it was concluded that the star is possibly an eclipsing binary. The star was
classified as an unsolved variable and named V4386\,Sgr. There were about a dozen or so deviating points in the ASAS-3 data,
clearly indicating the presence of a narrow eclipse. Fortunately, the ASAS-3 data are better distributed in time than those from Hipparcos,
allowing to unambiguously derive the orbital period.  The period is equal to 10.79824~d. The light curve (Fig.~\ref{fig3}) shows also a small
reflection effect.  The analysis of the out-of-eclipse data reavealed the presence of at least four pulsational components with frequencies
between 5.38 and 7.55~d$^{-1}$ (Tab.~\ref{bc-new-f}), indicating clearly that the primary component is a $\beta$~Cephei-type pulsator.
This is in agreement with
the published MK spectral types of this star: B0.5\,III \citep{hill74}, B1\,II \citep{garr77}, and B1\,Ib/II \citep{houk2}.

As can be seen in Fig.~\ref{fig3}, the eclipse covers only about 0.04 in phase, corresponding to about 10 hours, and is probably total. In the lower
panel of Fig.~\ref{fig3}, Hipparcos data obtained close to the primary minimum are also plotted. The five deviating points mentioned earlier
fit the eclipse quite well. However, due to the small number of data points, the contribution from pulsations could not be removed from the Hipparcos
data.  This may account for the deviations with respect to the ASAS-3 light curve (dots).

\subsubsection{HD\,168050}
\begin{figure}
\includegraphics[width=83mm]{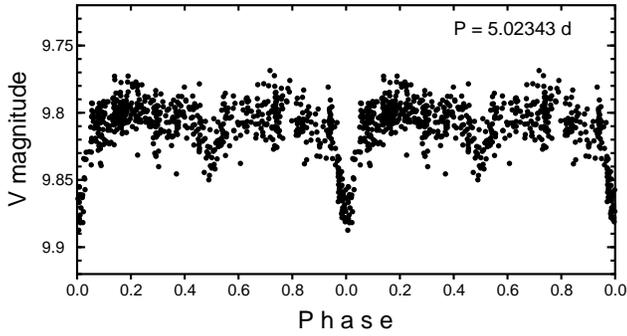}
 \caption{The ASAS-3 $V$-filter light curve of HD\,168050 freed from the contribution of two pulsating modes and folded
with the orbital period of 5.02343~d.}
\label{fig4}
\end{figure}

The fourth $\beta$~Cephei star we discovered in an eclipsing system is HD\,168050. From the
analysis of the ASAS-3 data we found two periodic terms with frequencies 5.549 and 5.251~d$^{-1}$, certainly
coming from pulsations. It is interesting to note
that the first mode shows large period increase (see Sect.~3.5), a phenomenon which is rather rare among $\beta$~Cephei stars.
The eclipsing light-curve of this star is shown in Fig.~\ref{fig4}. The orbital period is equal to 5.02335~d and the eclipses are rather shallow. The depth of the secondary minimum is about 1/3 of that of the primary one which
indicates that the secondary component is relatively massive. This also means
that the contribution of the secondary to the spectrum of the system is significant and may explain why the spectral type given by
\citet{houk4}, B3/5\,Ib, is slightly too late for a $\beta$~Cephei-type star.

\subsection{Distribution in Galactic coordinates}
\begin{figure*}
\sidecaption
\includegraphics[width=12cm]{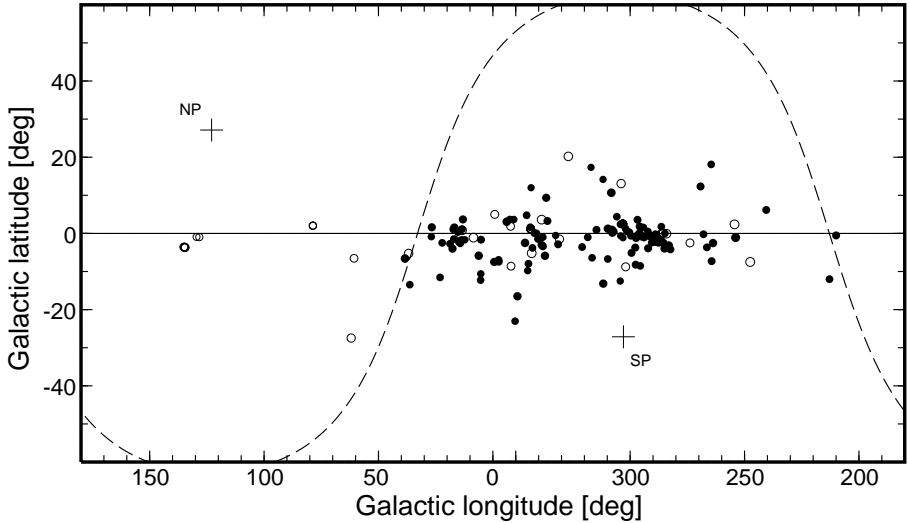}
\caption{Distribution of $\beta$~Cephei stars in Ga\-lac\-tic coordinates.  Only stars with $V$ magnitudes between 7 and 11~mag are shown.
Open circles denote stars from \citet{stha05}, filled circles, stars found in the ASAS-3 data.
`NP' and `SP' indicate the northern and southern celestial pole, respectively. 
Dashed line is the celestial equator.}
\label{gal-bc}
\end{figure*}

Since $\beta$~Cephei stars are very young, we expect to find them rather close to the Galactic plane. As discussed by
\citet{pigu05}, the known stars of this type are indeed located at low Galactic latitudes, except for a few nearby stars and an interesting
high-latitude star, HN\,Aqr \citep{waru88,lynn02}. A strip of stars belonging to the Gould Belt
could be also identified. However, the observed distribution was biased due to observational selection effects. The homogeinity of
the ASAS-3 data and the Michigan catalogues allows studying this distribution quantitatively. As can be seen in Fig.~\ref{gal-bc},
the new members of the $\beta$~Cephei family are distributed very unevenly in Galactic coordinates: they tend to form at least
three large groups. The largest group, with a center approximately at ($l$, $b$) = (300$^\circ$, 0$^\circ$) agrees with the direction in which we look
along the Sagittarius-Carina spiral arm. We postpone a detailed discussion of this distribution until a future paper, in which the data for the
remaining $\beta$~Cephei stars will be given. We just indicate here that there are several stars with quite large Galactic latitudes which, taking
into account their absolute magnitudes and low reddenings, are situated at large distances both from the Sun and from the
Galactic plane. Their distances might be comparable to the distance of the Galactic center, while the distance from the Galactic
plane might be as large as 1 kpc or even larger.

\subsection{Stars in open clusters and associations}
We have already indicated in Paper I that owing to the poor spatial resolution, the ASAS-3 data are not well suited for the detection
of $\beta$~Cephei stars in open clusters. However, if a cluster is sparse or the variable is far from the central area of the cluster,
the ASAS-3 photometry can allow detection. In the sample of the 103 new $\beta$~Cephei stars we can indicate six stars which are
members or likely members of young open clusters, and other six which are members of OB associations. This is indicated in the last
column of Tab.~\ref{bc-new}. The membership of new $\beta$~Cephei stars in clusters and their distribution in the Galaxy will be discussed in the next
paper of the series.

\subsection{Hipparcos stars}
The original work on Hipparcos data resulted in the discovery of six $\beta$~Cephei stars \citep{wael98,aert00}. Later on,
some follow-up work was done \citep{koen01,koey02} yielding about 60 candidates for short-period variables among O and early
B-type stars of which at least some are expected to be $\beta$~Cephei stars. Due to the unfavourable distribution of the Hipparcos
data in time, this discovery needs, however, to be confirmed by independent observations. With the ASAS-3 data,
we could verify the variability of 22 of these candidates; they are listed
in Tab.~\ref{hipp-ver}. The remaining stars are either saturated in the ASAS-3 data or are located north of declination $+$28$\degr$, i.e., outside the area in the sky
covered by the ASAS-3 observations.

\begin{table*}
 \caption{The results of the verification of variability of 22 O and early B-type candidate short-period variables observed by Hipparcos and found by 
\citet{koen01} and \citet{koey02}. DT is the detection threshold in the ASAS-3 photometry.}
  \begin{tabular}{crcrrrrrrrl}
  \hline\hline\noalign{\smallskip}
& & ASAS & \multicolumn{1}{c}{$V$} & \multicolumn{1}{c}{$f_{\rm Hipp}$} & \multicolumn{1}{c}{$A_{\rm Hipp}$} & \multicolumn{1}{c}{$f_{\rm ASAS}$} &
 \multicolumn{1}{c}{$A_{\rm ASAS}$} & S/N & DT & \\
HIP & HD(E) & name & [mag] & \multicolumn{1}{c}{[d$^{-1}$]} & \multicolumn{1}{c}{[mmag]} & \multicolumn{1}{c}{[d$^{-1}$]} & \multicolumn{1}{c}{[mmag]} & 
& [mmag] & Remarks\\
 \noalign{\smallskip}\hline\noalign{\smallskip}
29429 & 42597 & 061159$+$0723.5 & 7.04 & 10.67562 & 6.7 & --- & --- & --- & 8.0 & \\
30393 & 44597 & 062329$+$2023.5 & 9.04 & 3.25095 & 9.9 & --- & --- & --- & 7.3 &  \\
35355 & 56847 & 071810$-$1537.7 & 8.92 & 10.68006 & 26.4 & --- & --- & --- & 10.3 & MU CMa\\
36727 & 60308 & 073313$-$1527.2 & 8.20 & 11.20005 & 16.9 & --- & --- & --- & 9.1 &\\
38896 & 65658 & 075738$-$4635.6 & 7.23 & 11.76946 & 7.2 &  --- & --- & --- & 4.3 & \\
40268 & 68962 & 081323$-$3618.7 & 7.34 & 4.83740 & 7.6 & 4.76251 & 7.4 & 6.9 & 4.3 & new $\beta$~Cephei star\\
46876 & 82830 & 093309$-$4645.9 & 9.23 & 7.35793 & 16.9 & --- & --- & --- & 9.2 & \\
48652 & 86085 & 095519$-$3845.8 & 8.93 & 7.85499 & 11.6 & 7.87529 & 9.2 & 9.4 & 3.9 &new $\beta$~Cephei star\\
50598 & 89767 & 102012$-$5236.1 & 7.21 & 11.08927 & 12.6 & --- & --- & --- & 8.4 & V342 Vel\\
55499 & 306387 & 112159$-$6101.8 & 9.58 & 4.80836 & 31.6 & 4.80836 & 34.5 & 26.5 & 5.2 & new $\beta$~Cephei star,\\
&&&&&& 4.99402 & 14.6 & 11.2 &&V906\,Cen\\
62115 & 311884 & 124351$-$6305.2 & 10.82 & 11.41452 & 40.3 & 0.16028 & 52.9 & 19.8 & 10.7 & CD Cru (E), W-R star\\
63489 & 112842 & 130031$-$6022.5 & 7.07 & 11.26044 & 8.4 &  --- & --- & --- & 7.1 & \\
67060 & 119608 & 134431$-$1756.2 & 7.50 & 10.55676 &13.8 & --- & --- & --- & 7.9 &\\
69847 & 124788 & 141742$-$6047.3 & 8.98 & 0.45821 & 52.8 &  0.22917 & 70.5 & 97.8 & 5.6 &  V999 Cen (E or Ell)\\
 &&&& 9.13446 & 21.0 & --- & --- & --- & &\\
70228 & 125545 & 142204$-$5817.4 & 7.42 & 11.50208 & 14.9 & --- & --- & --- & 7.7 &\\
76642 & 137179 & 153903$-$8314.0 & 8.75 & 10.08788 & 13.0 & --- & --- & --- & 4.0 &\\
80408 & 147421 & 162445$-$5327.8 & 9.02 & 5.35481 & 25.1 & 5.35473 & 27.1 & 19.0 & 2.8 & new $\beta$~Cephei star\\
81216 & 149100 & 163521$-$5338.9 & 7.20 & 5.16114 & 6.0 & 5.16118 & 8.1 & 7.7 & 3.8 & new $\beta$~Cephei star\\
83603 & 154043 & 170519$-$4704.1 & 7.08 & 11.17594 & 15.8 & --- & --- & --- & 9.7 & V863 Ara\\
91822 & 173003 & 184318$-$0138.7 & 7.68 & 7.98590 & 9.1 & --- & --- & --- & 3.2 &\\
98807 & 339483 & 200401$+$2616.3 & 8.94 & 3.72410 & 23.4 &3.72413 & 22.5 & 14.0 & 6.4 & new $\beta$~Cephei star \\
99327 & 191531 & 200940$+$2104.7 & 8.35 & 6.08588 & 26.5 &  6.08589 & 26.0 & 12.2 & 8.6 & $\beta$~Cephei \citep{hand05} \\
\noalign{\smallskip}\hline
\end{tabular}
\label{hipp-ver}
\end{table*}

Of these 22 stars, only seven can be classified as $\beta$~Cephei stars. One, HD\,191531, was already discovered by \citet{hand05}, see also Paper I.
The remaining six are new $\beta$~Cephei stars, but only four are listed in Tab.~\ref{bc-new}, while two,
HDE\,306387 = V909\,Cen and HDE\,339483, are not included because they do not appear in the Michigan catalogues.
These two stars will be discussed in the next paper of the series.

For the remaining 15 stars from Tab.~\ref{hipp-ver}, the short-period variability was not confirmed by the ASAS-3 data, although for all but HD\,42597 the
detection threshold was smaller than the amplitude reported from the analysis of the Hipparcos data. It can be seen from Tab.~\ref{hipp-ver} that frequencies
for most of them range between 10 and 12~d$^{-1}$. As already explained by \citet{jepa00} and \citet{koey02}, these spurious frequencies
are due to aliasing originating from the convolution of the low-frequency changes and the spectral window dominated by the rotation frequency of the
satellite (11.25~d$^{-1}$).

For five of the seven $\beta$~Cephei stars in Tab.~\ref{hipp-ver}, the agreement between frequencies derived from the Hipparcos and ASAS-3 data
is very good. Only for HD\,68962 and 86065 the differences are significant, amounting to about $\Delta f$ = $f_{\rm ASAS} - f_{\rm Hipp}$ =
$-$0.0749 and $+$0.0203~d$^{-1}$, respectively. This can be explained in terms of aliasing in the periodograms of Hipparcos data.

\section{Discussion and conclusions}
According to theory \citep[see, e.g.,][]{pamy99}, for the zero-age main-sequence models of massive stars, acoustic (p) and gravity (g)
modes separate very well in frequency. This separation is not so good for evolved stars because in the course of the main-sequence evolution
frequencies of g modes increase and eventually replace those of p modes through avoided crossing \citep{dzpa93}. Such modes have a `mixed' character:
of a p mode in the envelope and of a g mode in the stellar interior.
When only unstable modes are considered, however, the separation in frequency of p and g modes is, for a given model, preserved. Pulsations in p modes 
and mixed modes are attributed to $\beta$~Cephei-type variability, whereas those in g modes, to the SPB type. As $\beta$~Cephei stars
populate mainly the later stages of main-sequence evolution, mixed modes are expected to be excited in these stars.

Theoretical calculations show also \citep{pamy99} that $\beta$~Cephei and SPB instability strips partially overlap, which means that hybrid stars,
showing pulsations of both types, might exist. This possibility was already indicated by \citet{dzie93} for a region around spectral type B2-B3 and masses
equal to 7--8~$M_\odot$. Newer calculations \citep{pamy99} confirmed this and showed in addition that g-mode instability extends towards higher masses.
The latter instability is confined, however, to the very late main-sequence stages. 

This picture of p and g mode instability domains was calculated neglecting the effects of rotation on pulsations. The theory that includes fast
rotation is still not fully satisfactory, but it is already clear that rotation changes stability of modes. A good example of the theoretical calculations that include the
effect of rotation on the stability of g modes was published by \citet{town05}, who showed that for fast-rotating stars the SPB
instability strip widens towards higher masses. The region of possible hybrid
behaviour becomes therefore much larger.  In addition, the frequencies of modes, especially of non-axisymmetric modes, change considerably.
In particular, frequencies of retrograde sectoral mixed modes are shifted to much lower frequencies than they would
have in non-rotating stars, i.e., will occur in the range where normally the g modes are observed. On the other hand, frequencies
of prograde sectoral g modes shift to the region occupied normally by p modes.
It must be also remembered that in the presence of fast rotation modes can no longer be classified in a simple scheme described by three quantum numbers
because the geometry of such modes differs from spherical harmonics.

The effects related to fast rotation are impontant not only because some $\beta$~Cephei stars are known to
rotate fast. The region of $\beta$~Cephei and SPB stars in the H-R diagram is populated by Be stars which,
by definition, rotate very fast. These stars were known to
show both short-period variability and line-profile variations. The short-period photometric variability called, after the prototype, the
$\lambda$~Eri-type \citep{balo90} was
usually attributed to rotationally-induced effects because no clear evidence for multiperiodicity was found \citep[see, e.g.,][]{baja02}. On the other hand, there
were many convincing examples of line-profile variations in Be stars that were interpreted in terms of non-radial pulsations 
\citep{baad84,yang90,stef95,floq00}, including
multiperiodic behaviour \citep{rivi98}. Despite this, it was not widely accepted that non-radial pulsations are present in Be stars mainly because
there was no clear relation between periods seen in photometry and spectroscopy. An excellent review of variabilities observed in Be stars
was recently given by \citet{pori03}.

A strong argument in favour of non-radial pulsations in Be stars, as seen in photometry, came from the observations carried out by the MOST satellite,
in particular, those of an early Be star, HD\,163868, in which \citet{walk05-Be} found over 60 periodicities in three distinct groups. A theoretical interpretation of
this variability \citep{walk05-Be,dzie07,savo07} includes pulsations in g modes. It is therefore reasonable to use the term `SPBe stars' for
Be stars with detected g-mode pulsations, as proposed by \citet{walk05-Be}, although it has to be remembered that many previous investigators
already indicated a common cause of periodic variability in SPB and Be stars.

The two grouping of modes seen in HD\,163868 around 1.7 and 3.4 d$^{-1}$ might be a clue to understanding a grouping of modes
in $\beta$~Cephei and SPB candidates in the LMC and SMC, found by \citet{kola04,kola06}. In many of these stars, the frequencies of
modes grouped around two values whose ratio was close to 2.
Whether fast rotation is responsible for this could be verified by future spectroscopic observations.
For the SMC we know that many of these stars are fast-rotating because they show emission in Balmer lines.

Since inclusion of fast rotation enhances the possibility of getting unstable low-frequency modes among early and mid-B stars, we should see them observationally.
It seems that we get more and more examples that confirm these predictions. First, low-frequency modes were detected in two classical $\beta$~Cephei stars,
$\nu$~Eri \citep{jerz05} and 12~Lac \citep{hand06}. In the ASAS-3 data (Sect.~3.4), we found five $\beta$~Cephei stars in which low-frequency
modes were detected. In addition, another $\beta$~Cephei star with low-frequency modes, HD\,133823, was found in Paper I
(see also Fig.~\ref{fdf}). On the other hand, periodic variations with very short periods were found in some mid and late B-type stars like
V2104\,Cyg \citep{uytt07} and stars in NGC\,3293 \citep{hand07}. A preliminary analysis of the data obtained for the other young open cluster,
h Persei (NGC\,869) also reveals the presence of mid-B stars, of which some are known as Be stars, showing relatively short periods, even below 0.3~d
(Majewska-\'Swierzbinowicz et al., in preparation).

The complexity of mode behaviour in the presence of fast rotation may
pose a problem as far as the classification of hot pulsators is concerned. This is because period(s) alone may not be sufficient to
recognize $\beta$~Cephei and SPB star.  Mode identification becomes a neccessity if we want to classify stars pulsating in p modes as $\beta$~Cephei stars, while those pulsating in g modes, as SPB stars.

The other important results that come from the discovery of 103 $\beta$~Cephei stars can be summarized as follows:
\begin{enumerate}
 \item The number of known $\beta$~Cephei stars has grown to over 200 and thus some statistical studies become possible. 
The sample of
new stars is particularly well suited for this purpose because the stars were selected and observed in a homogeneous way.
\item From the new sample, a selection of excellent targets for asteroseismology can be done. These would include stars
with large amplitudes (Sect.~3.2), large number of modes, equidistant multiplets (Sect.~3.3) and components of
eclipsing binaries (Sect.~3.7).
\item We have found three O-type $\beta$~Cephei stars (Sect.~3.6); this type of pulsation was
also found in $\zeta$~Oph from the MOST data \citep{walk05}.  Thus, we may conclude that $\beta$~Cephei-type instability extends
towards O-type stars, but seems to be confined to very late subtypes (O8-09). This still does not agree well with the theory because
calculations do not predict an upper luminosity limit for p-mode instability \citep{pamy99}. However, these calculations have not included the effect of stellar wind which may stabilize pulsations for hotter O-type stars.
\item
The very recent stability calculations using new opacities \citep{migl07} indicate a much wider region in
the H-R diagram where hybrid, $\beta$~Cephei/SPB-type behaviour, should be observed.
The inclusion of fast rotation seems to have a similar effect \citep{town05}.  Observationally, this fact is confirmed by the growing
number of detections of hybrid stars, including five $\beta$~Cephei stars with long periods we found (Sect.~3.4) as candidates.
In order to verify their hybridity, mode identification is needed. A detailed study of these stars may allow to testing the physics used in calculations of the evolutionary models of massive stars but also in the theoretical
predictions of their pulsational stability.
\end{enumerate}
Finally, it is interesting to note that four stars from our sample, HD\,48553, 171305, 173006, and 186610 are located within the `CoRoT eyes' \citep{bagl02} and,
if selected for observations by this satellite, could be
studied in unprecedented detail.

\begin{acknowledgements}
The work was supported by the MNiI/MNiSzW grants No. 1 P03D 016 27 and N203 007 31/1328. We greatly acknowledge comments made by Prof.~M.\,Jerzykiewicz and Dr.~Gerald Handler, the referee. This research has made use of the SIMBAD database,
operated at CDS, Strasbourg, France.
\end{acknowledgements}

\setcounter{table}{0}
{\small
\begin{longtable}{rcccrrrll}
\caption{\label{bc-new}New $\beta$~Cephei stars.  $N$ is the number of independent modes found in the ASAS-3 photometry,
the remaining columns are self-explanatory. References to the MK spectral types in the eighth column are the following:
(1) \citet{houk1}; (2) \citet{houk2}; (3) \citet{houk3}; (4) \citet{houk4}; (5) \citet{houk5}; (6) \citet{hill70}.}\\
\hline \hline
& & & & \multicolumn{1}{c}{$V$} & $B-V$ & $U-B$ & &  \\
HD & CPD/BD & ASAS name & $N$ & [mag] & [mag] & [mag] & MK sp.~type& Notes \\
\hline
\endfirsthead
\caption{continued.}\\
\hline
\endhead
\hline
\endfoot
 46994 & $-$25$\degr$1525 & 063513$-$2550.4 & 1 & 7.86 & $-$0.17 & ---   & B2/3 V (4) & in CMa OB2\\
 48553 & $+$02$\degr$1369 & 064410$+$0223.5 & 1 & 9.03 & --- & --- & B2 III (5) & in the CoRoT field\\
 67600 & $-$20$\degr$3392 & 080754$-$2109.0 & 1 & 9.18 & --- & --- & B3/5 V (4) &\\
 68962 & $-$35$\degr$2054 & 081323$-$3618.7 & 1 & 7.34 & $+$0.15 & $-$0.65 & B2/3 V (3) &in BH 23\\
 69016 & $-$27$\degr$2973 & 081401$-$2743.1 & 1 & 10.22 & --- & --- & B3 IV (3) &\\
 69824 & $-$48$\degr$1585 & 081638$-$4826.2 & 1 & 9.10 & --- & --- & B4/6 V (2) &\\
 73568 & $-$44$\degr$2824 & 083719$-$4512.5 & 2 & 8.35 & $+$0.30 & $-$0.57 & B2/3 IV (2)& in Vela OB1\\
 74339 & $-$47$\degr$2545 & 084133$-$4801.5 & 6 & 9.30 & $+$0.14 & $-$0.67 & B2/3 II/III (2) & in IC 2395\\
 77769 & $-$46$\degr$3350 & 090248$-$4657.9 & 1 & 9.38 & --- & --- & B3 IV (2) & in Platais 9\\
 86085 & $-$38$\degr$3738 & 095519$-$3845.8 & 1 & 8.94 & $-$0.09 & $-$0.85 & B1 Vn (6) &halo star\\
\noalign{\medskip}
 86214 & $-$59$\degr$1528 & 095458$-$5949.8 & 6 & 9.21 & $-$0.02 & $-$0.79 & B2 Ib (1)&\\
 86248 & $-$30$\degr$3001 & 095633$-$3126.5 & 1 & 9.63 & $-$0.19 & $-$0.89 & B2 III (3) &high-velocity star, very distant\\
 87592 & $-$58$\degr$1860 & 100409$-$5922.9 & 5 & 8.96 & --- & --- & B1 IV (1) & \\
 88844 & $-$60$\degr$1765 & 101302$-$6110.7 & 2 & 8.53 & $-$0.16 & $-$0.93 & B0.5 III (1) & ALS 1453\\
 90075 & $-$59$\degr$2079 & 102206$-$6011.2 & 1 & 8.88 & $+$0.07 & $-$0.76 & B1 II (1) & ALS 1517\\
 90987 & $-$57$\degr$3308 & 102849$-$5746.0 & 4 & 9.64 & $+$0.13 & $-$0.70 & B1/2 III/IVp (1) & ALS 1601, in IC 2581 ?\\
 91651 & $-$59$\degr$2214 & 103330$-$6007.7 & 2 & 8.85 & $-$0.01 & $-$0.89 & O8/9 V (1) & ALS 1647, in Car OB1\\
 92291 & $-$60$\degr$2077 & 103756$-$6116.5 & 5 & 9.07 & $-$0.07 & $-$0.86 & B2 Ib/II (1) & ALS 1702\\
 93113 & $-$57$\degr$3449 & 104400$-$5741.1 & 1 & 8.71 & $+$0.03 &$-$0.76 & B1/2 II/III (1) & ALS 1818\\
 93341 & $-$56$\degr$3759 & 104524$-$5659.4 & 1 & 9.64 & $+$0.02 & $-$0.74 & B1/2 II/III (1) & ALS 1876\\
\noalign{\medskip}
 94065 & $-$61$\degr$1905 & 105007$-$6148.3 & 5 & 8.95 & $-$0.05 & $-$0.83 & B0/1 III-IV (1) & \\
 94345 & $-$59$\degr$2802 & 105210$-$5957.3 & 3 & 9.00 & $+$0.06 & $-$0.77 & B0/1 IV (1) & \\
 94900 & $-$59$\degr$2859 & 105611$-$5955.8 & 1 & 9.25 & $+$0.17 & $-$0.66 & B1/2 III (1) & ALS 2034\\
 95568 & $-$61$\degr$2021 & 110040$-$6236.9 & 2 & 9.56 & $+$0.12 & --- & O9/B0 V (1) & ALS 2084\\
 96882 & $-$60$\degr$2553 & 110824$-$6117.1 & 1 & 9.03 & $-$0.03 & $-$0.86 & B1 II/III (1) &\\
 96901 & $-$63$\degr$1845 & 110820$-$6436.1 & 2 & 8.87 & $+$0.20 & $-$0.62 & B1 III/IV (1) &\\
 97629 & $-$60$\degr$2677 & 111308$-$6106.8 & 1 & 9.53 & $+$0.11 & $-$0.65 & B3/5 Ib/II (1) &  ALS 2250\\
 98260 & $-$60$\degr$2783 & 111719$-$6130.7 & 3 & 9.70 & $+$0.26 & $-$0.63 & B1/2 Ib/II (1) & ALS 2292\\
 99024 & $-$59$\degr$3411 & 112303$-$6031.2 & 1 & 9.66 & $+$0.03 & $-$0.77 & B2 Ib/II (1) & ALS 2327\\
 99205 & $-$69$\degr$1537 & 112356$-$7008.0 & 2 & 9.59 & $+$0.01 & $-$0.73 & B1 III (1) & ALS 2336\\
\noalign{\medskip}
100355 & $-$61$\degr$2391 & 113216$-$6232.8 & 3 & 9.23 & 0.00 & $-$0.67 & B2/3 III (1) &\\
101794 & $-$61$\degr$2541 & 114225$-$6228.6 & 2 & 8.68 & $+$0.06 & $-$0.74 & B0/1 n (1) & ALS 2460, eclipsing, Be star,\\
&&&&&&&&in Stock 14\\
101838 & $-$61$\degr$2550 & 114249$-$6233.9 & 1 & 8.42 & $+$0.02 & $-$0.81 & B0.5/1 III (1) & ALS 2463, eclipsing, in Stock 14\\
102505 & $-$69$\degr$1588 & 114720$-$7025.0 & 2 & 9.02 & $+$0.08 & --- & B1 IV (1) & ALS 2495\\
103007 & $-$61$\degr$2696 & 115118$-$6159.8 & 3 & 9.29 & $-$0.01 & $-$0.82 & B2/3 (III) (1)& ALS 2519\\
103320 & $-$59$\degr$3910 & 115341$-$6016.0 & 1 & 8.82 & $-$0.03 & --- & B2 III (1) & \\
103764 & $-$61$\degr$2814 & 115652$-$6236.6 & 1 & 9.48 & $-$0.06 & $-$0.80 & B2 II/III (1) &\\
104257 & $-$65$\degr$1757 & 120012$-$6604.7 & 1 & 8.85 & $+$0.05 & $-$0.73 & B1/2 II/III (1) & ALS 2562\\
104465 & $-$62$\degr$2526 & 120137$-$6333.8 & 2 & 9.07 & $-$0.05 & $-$0.81 & B2 II/III (1) & \\
104795 & $-$58$\degr$4048 & 120400$-$5843.4 & 1 & 9.09 & $+$0.11 & --- & B2 III (1) &\\
\noalign{\medskip}
106345 & $-$67$\degr$1921 & 121416$-$6744.4 & 1 & 9.08 & --- & --- & B2 III (1) &\\
108628 & $-$61$\degr$3220 & 122906$-$6228.0 & 2 & 9.16 & --- & --- & B2 II (1) &\\
108769 & $-$33$\degr$3261 & 123003$-$3429.9 & 1 & 9.03 & --- & --- & B2 III (3) & \\
110498 & $-$60$\degr$4253 & 124316$-$6138.9 & 1 & 9.67 & $+$0.49 & $-$0.49 & B0.5 III (1) & ALS 2737, in Cen OB1\\
111377 & $-$60$\degr$4309 & 124938$-$6057.9 & 1 & 9.52 & $+$0.13 & $-$0.68 & B2/5 II-III (1) & ALS 2775\\
111578 & $-$59$\degr$4488 & 125057$-$6026.8 & 1 & 9.13 & $+$0.16 & $-$0.57 & B2/3 II (1) & ALS 2785\\
113013 & $-$59$\degr$4653 & 130141$-$6026.4 & 2 & 9.40 & $+$0.12 & --- & B2 III (1) & ALS 2871\\
114444 & $-$74$\degr$1029 & 131304$-$7518.9 & 2 & 10.32 & $-$0.02 & $-$0.80 & B1 II/III (1) & ALS 2963\\
114733 & $-$57$\degr$5952 & 131330$-$5821.7 & 1 & 9.52 & $+$0.27 & $-$0.55 & B2/3 III (1) & ALS 2980\\
115533 & $-$61$\degr$3586 & 131901$-$6153.7 & 1 & 10.04 & $+$0.08 & $-$0.76 & B2/5 (Ib/II) (1) & ALS 3026, Vis.~double, sep.~6\arcsec  \\
\noalign{\medskip}
116538 & $-$51$\degr$6048 & 132512$-$5150.5 & 1 & 7.90 & $-$0.06 & $-$0.89 & B2 Ib/II (2) &  \\
116827 & $-$61$\degr$3706 & 132735$-$6207.1 & 2 & 9.51 & $+$0.11 & $-$0.57 & B2/5 (Vn) (1) & Be star\\
117357 & $-$61$\degr$3760 & 133116$-$6144.0 & 2 & 9.02 & $+$0.23 & $-$0.72 & O9.5/B0 V (1) & ALS 3103, Be star\\
117687 & $-$60$\degr$4759 & 133330$-$6126.8 & 1 & 9.33 & $+$0.12 & $-$0.71 & B2 Ib/II (1) & ALS 3115\\
117704 & $-$61$\degr$3793 & 133337$-$6219.1 & 1 & 8.89 & $+$0.20 & $-$0.59 & B1/2 III (1) & ALS 3116, close to WR 55\\
119252 & $-$47$\degr$6162 & 134309$-$4748.3 & 1 & 10.34 & --- & --- & B3 III (2) &\\
119910 & $-$60$\degr$4978 & 134754$-$6053.0 & 2 & 8.31 & $-$0.07 & --- & B2 III/IV (1) &\\
122831 & $-$67$\degr$2491 & 140701$-$6834.1 & 2 & 9.60 & $+$0.05 & --- & B1 III (1) & ALS 3196\\
123077 & $-$42$\degr$6537 & 140637$-$4327.8 & 4 & 10.30 & $-$0.23 & --- & B2/3 II (2) &\\
126357 & $-$59$\degr$5569 & 142709$-$5944.2 & 3 & 9.07 & $+$0.22 & $-$0.64 & B1 III (1) & ALS 3238, in NGC 5606\\
\noalign{\smallskip}
131805 & $-$73$\degr$1416 & 150055$-$7347.7 & 1 & 8.47 & --- & --- & B5 V (1) &long-term variability superimposed\\
132320 & $-$59$\degr$5775 & 150106$-$5953.7 & 1 & 9.40 & $+$0.31 & $-$0.53 & B1 III (1) &\\
137405 & $-$60$\degr$5814 & 152854$-$6048.1 & 1 & 9.36 & $+$0.13 & $-$0.64 & B0.5/1 Iab (1)& ALS 3356\\
142754 & $-$40$\degr$7155 & 155749$-$4059.8 & 1 & 8.60 & $+$0.17 & $-$0.60 & B2 III (2) &\\
145537 & $-$34$\degr$6468 & 161255$-$3439.3 & 2 & 10.41 & $+$0.10 & $-$0.78 & B2 Ib/II (3)&\\
146442 & $-$45$\degr$7845 & 161823$-$4550.4 & 3 & 9.04 & $+$0.30 & $-$0.51 & B2 III (2)&\\
147421 & $-$53$\degr$7928 & 162445$-$5327.8 & 1 & 9.02 & $+$0.09 & $-$0.81 & B0 II (1)& ALS 3596\\
149100 & $-$53$\degr$8064 & 163521$-$5338.9 & 1 & 7.19 & $-$0.06 & $-$0.76 & B2 III (1) &\\
150927 & $-$37$\degr$6738 & 164544$-$3810.1 & 2 & 9.44 & $+$0.32 & $-$0.52 & B2/3 Ib (3)&\\
151158 & $-$42$\degr$7512 & 164734$-$4252.7 & 1 & 8.20 & $+$0.22 & $-$0.49 & B2 Ib/II (2)& ALS 3754\\
\noalign{\medskip}
152060 & $-$41$\degr$7683 & 165300$-$4124.5 & 2 & 9.58 & $+$0.06 & $-$0.73 & B1 III (2) &\\
152162 & $-$45$\degr$8186 & 165354$-$4600.9 & 2 & 8.59 & $+$0.18 & --- & B2 III (2) &\\
152372 & $-$48$\degr$8913 & 165517$-$4848.3 & 3 & 8.85 & $+$0.20 & $-$0.60 & B2 III/IV (2)&\\
153772 & $-$50$\degr$9813 & 170400$-$5105.0 & 1 & 8.32 & $+$0.06 & $-$0.61 & B2 V (2)& in R assoc.~vdBH81a\\
154500 & $-$34$\degr$6724 & 170721$-$3428.7 & 2 & 9.94 & $+$0.22 & $-$0.56 & B2 Iab/b (3) &\\
155407 & $-$45$\degr$8378 & 171332$-$4527.2 & 1 & 9.52 & --- & --- & B2 II (2) &\\
156172 & $-$41$\degr$7933 & 171800$-$4203.6 & 1 & 8.18 & $+$0.40 & $-$0.58 & O9.5/B0.5 Ia/Iab (2) & ALS 4037\\
156321 & $-$32$\degr$4462 & 171820$-$3219.6 & 1 & 8.15 & $+$0.01 & $-$0.77 & B2 V (3)&\\
159792 & $-$46$\degr$8786 & 173850$-$4618.9 & 3 & 9.43 & $+$0.15 & $-$0.63 & B2 II (2)&\\
161633 & $-$46$\degr$8909 & 174900$-$4656.4 & 4 & 9.94 & $+$0.12 & $-$0.95 & B1 Iab/b (2)&\\
\noalign{\medskip}
164188 & $-$15$\degr$4767 & 180039$-$1548.1 & 1 & 8.67 & $+$0.19 & $-$0.71 & B1 Ib/II (4)& ALS 4546\\
164741 & $-$25$\degr$6269 & 180344$-$2518.7 & 5 & 9.03 & $+$0.31 & $-$0.55 & B2 Ib/II (4) & ALS 4590\\
165955 & $-$34$\degr$7625 & 180958$-$3452.1 & 2 & 9.20 & $-$0.05 & $-$0.82 & B2 II/III (3) & ALS 4702, high-velocity star\\
166304 & $-$16$\degr$4739 & 181043$-$1642.8 & 3 & 9.33 & $+$0.19 & $-$0.63 & B2 II (4) & \\
167003 & $-$33$\degr$4963 & 181442$-$3308.5 & 4 & 8.45 & $-$0.13 & $-$0.94 & B1 Ib/II (3) & ALS 4801, V4386 Sgr, eclipsing\\
167451 & $-$13$\degr$4897 & 181555$-$1334.5 & 1 & 8.24 & $+$0.78 & $-$0.29 & B1 Ib (4) & ALS 4844, in Ser OB1\\
168015 & $-$13$\degr$4919 & 181822$-$1323.2 & 2 & 9.11 & $+$0.36 & --- & B3/5 II (4) & ALS 9488\\
168050 & $-$19$\degr$6762 & 181839$-$1906.2 & 2 & 9.81 & --- & --- & B3/5 Ib (4) & eclipsing\\
168675 & $-$17$\degr$5154 & 182143$-$1753.8 & 2 & 8.94 & $+$0.28 & --- & B2 Ib/II (4) & ALS 4964\\
168750 & $-$26$\degr$6414 & 182221$-$2625.0 & 3 & 8.27 & $+$0.08 & $-$0.74 & B1 Ib (3) & ALS 4969\\
\noalign{\medskip}
169173 & $-$17$\degr$5179 & 182408$-$1751.8 & 1 & 9.93 & --- & --- & B3 Ib (4) &\\
169601 & $-$17$\degr$5192 & 182607$-$1747.3 & 3 & 9.88 & --- & --- & B0/1 Iab/b (4) &\\
171141 & $-$46$\degr$9389 & 183516$-$4556.4 & 1 & 8.37 & $-$0.22 & $-$0.97 & B2 II/III (2) & halo star\\
171305 & $-$04$\degr$4497 & 183416$-$0448.8 & 1 & 8.70 & $+$0.45 & --- & B2 II (5) & ALS 9749, in the CoRoT field\\
171344 & $-$13$\degr$5039 & 183457$-$1352.2 & 1 & 9.52 & $+$0.29 & --- & B2 Ia (4) & ALS 9758\\
172140 & $-$29$\degr$5634 & 183948$-$2920.4 & 5 & 9.95 & $-$0.05 & $-$0.89 & B1 II (3) &\\
172427 & $-$10$\degr$4749 & 184032$-$1043.1 & 1 & 9.46 & $+$0.48 & $-$0.44 & B0.5/1 Ib (5) & ALS 9832, in Sct OB2\\
173006 & $-$05$\degr$4737 & 184326$-$0546.8 & 1 & 10.03 & $+$0.27 & $-$0.55 & B0.5 IV (5) & ALS 9902, in the CoRoT field\\
173502 & $-$30$\degr$5678 & 184656$-$2957.6 & 4 & 9.73 & $-$0.10 & $-$0.91 & B1/2 Ib (3) & halo star\\
178987 & $-$47$\degr$9208 & 191259$-$4709.7 & 1 & 9.83 & --- & --- & B2 II (2) &\\
\noalign{\medskip}
180032 & $-$14$\degr$5352 & 191531$-$1355.1 & 2 & 9.49 & $-$0.07 & --- & B2 III/IV (4) & ALS 10219\\
186610 & $-$03$\degr$4698 & 194527$-$0309.1 & 4 & 9.67 & $-$0.01 & $-$0.77 & B2 Ib/II (5) & ALS 10480, in the CoRoT field\\
187536 & $-$28$\degr$7032 & 195125$-$2813.5 & 1 & 9.46 & $-$0.08 & $-$0.86 & B3 II (3) &\\
\hline
\end{longtable}
}

\Online

\end{document}